\documentclass[12pt,preprint]{aastex}

\newcommand\mo{M$_\odot$}
\newcommand\col{\mbox {\rm cm$^{-2}$} }
\newcommand\kms{\mbox {km~s$^{-1}$} }
\newcommand\cxo{{\it Chandra\ }}
\newcommand\etal{et al.\ }
\begin{document}

\title{The Diffuse and Compact X-ray Components
of the Starburst Galaxy Henize~2-10 }

\shorttitle{X-rays in Henize~2-10}
\shortauthors{Martin \& Kobulnicky}

\author{Henry A. Kobulnicky}
\affil{Department of Physics \& Astronomy \\
	University of Wyoming \\
	Laramie, WY 82070}
\email{chipk@uwyo.edu}

\author{Crystal L. Martin}
\affil{University of California, Santa Barbara \\
	Department of Physics \\
	Santa Barbara, CA 93106}
\email{cmartin@physics.ucsb.edu}


\begin{abstract}
{\it Chandra X-ray Observatory} imaging spectroscopy of the starburst
galaxy Henize~2-10 reveals a strong nuclear point source and at least
two fainter compact sources embedded within a more luminous diffuse
thermal component.  Spectral fits to the nuclear X-ray source imply an
unabsorbed X-ray luminosity $L_x>10^{40}$ erg~s$^{-1}$ for reasonable
power law or blackbody models, consistent with accretion onto a $>$50
\mo\ black hole behind a foreground absorbing column of $N_H>10^{23}$
\col.  Two of these point sources have $L_x=2-5\times10^{38}$ erg
s$^{-1}$, comparable to luminous X-ray binaries.  These compact
sources constitute a small fraction ($\leq$16\%) of the total X-ray
flux from He~2-10 in the 0.3--6.0 keV band and just 31\% of the X-rays
in the hard 1.1--6.0 keV band which is dominated by diffuse emission.
Two-temperature solar-composition plasmas (kT$\simeq$0.2 keV and
kT$\simeq$0.7 keV) fit the diffuse X-ray component as well as
single-temperature plasmas with enhanced $\alpha$/Fe ratios.  Since
the observed radial gradient of the X-ray surface brightness closely
follows that of the H$\alpha$ emission, the composition of the X-ray
plasma likely reflects mixing of the ambient cool/warm ISM with an
even hotter, low emission measure plasma, thereby explaining the
$\sim$solar ISM composition. Aperture synthesis 21-cm maps show an
extended neutral medium to radii of 60\arcsec\ so that the warm and
hot phases of the ISM, which extend to $\sim$30\arcsec, are enveloped
within the 8$\times10^{20}$ cm$^{-2}$ contour of the cool neutral
medium.  This extended neutral halo may serve to inhibit a
starburst-driven outflow unless it is predominantly along the line of
sight.  The high areal density of star formation can also be reconciled with
the lack of prominent outflow signatures if Henize 2-10 is in the very
early stages of developing a galactic wind.

\end{abstract}

\subjectheadings{galaxies: formation --- galaxies: evolution ---
galaxies: fundamental parameters --- galaxies: abundances }

\section{Introduction}
\subsection{Goals of this Study}

The blue compact galaxy Henize~2-10 has been extensively studied as
the prototype of starbursting galaxies containing large populations of
Wolf-Rayet stars \citep{conti1991}.  As a galaxy hosting recent
intense star formation activity, He~2-10 is a natural target for
investigating the impact of starbursts on interstellar medium.  In
Henize 2-10, supernova activity has apparently produced a hot X-ray
emitting component and has arguably impacted the composition and
kinematics of the warm and cold ISM components as well
\citep{mendez1999, johnson2000, ott2005, grimes2005, schwartz2006}.
Our goal in this paper is to provide a systematic overview of the
physical state of and the relation between these components using a
panchromatic suite of X-ray, optical, and radio datasets. We use
a 20 ksec imaging observation with the {\it Chandra X-ray
Observatory} in order to characterize the distribution of the hot
X-ray emitting plasma in relation to the ionized and neutral gas
and provide a complete analysis of the X-ray point source population
in and around He~2-10.  Previous analyses of these {\it Chandra} data
by \citet{ott2005} and \citet{grimes2005} have addressed the
integrated X-ray properties of He~2-10 in the context of larger
samples of star-forming galaxies with diffuse X-ray emission.  These
studies concluded that the hot ISM is mildly enriched in
$\alpha$-process elements, consistent with the mixing of fresh
supernova ejecta from the recent starburst into the surrounding ISM.
Our focus here is a more detailed look at both the diffuse and compact
X-ray components of He~2-10 than previously reported.  We show that
$\alpha$-element enrichment is not strictly required to fit the X-ray
spectrum, we identify a hard ultraluminous X-ray source as a
counterpart to a nuclear radio source, and we discuss the relative
energetics and morphologies of the cold, warm, and hot phases of the
ISM to understand the state of the starburst-driven outflow.  First,
we provide a brief review of the global characteristics of He~2-10
that will be relevant to later discussion.

\subsection{Henize~2-10: Content and Kinematics }

Morphological studies reveal at least three distinct starforming
regions as shown in the images of \citet{mendez1999}.  The most
luminous central starburst region contains 6--10 compact ($<$1\arcsec)
super starclusters which have diameters of $<$10~pc at the nominal 9.8
Mpc distance of He~2-10.\footnote{The observed optical and 21-cm
radial velocities \citep{kobulnicky1995, mendez1999, schwartz2005}
corrected for Galactic rotation and Virgocentric infall yield a
Hubble-flow velocity of $\sim$712~\kms, leading to a distance of 9.8
Mpc for H$_0=72$ \kms\ Mpc$^{-1}$.  This implies an angular scale of
47~pc~arcsec$^{-1}$ and a distance modulus of 29.95 mag.  Throughout,
we convert linear distances and implied luminosities from prior works
self-consistently to this distance.  Distances ranging from 9 to 14
Mpc may be found in the literature.}  These clusters have ages $<$10
Myr and masses up to 10$^5$ M$_\odot$ \citep{contivacca1994,
johnson2000, vacca2002}. \citet{chandar03} assign an age range of 4--5
Myr for the four most luminous clusters.

Radio interferometer observations reveal the presence of at least four
radio-bright ``ultra-dense'' \ion{H}{2} regions powered by very young
super starclusters \citep{kj99, jk2003}.  The $H\alpha$ luminosity of
Henize~2-10 (not corrected for extinction) measured by
\citet{mendez1999} is $2.1\times10^{41}$ erg s$^{-1}$, leading to an
implied current star formation rate of $>$1.8 \mo\ yr$^{-1}$ based on
the formulation of \citet{kennicutt1989}.  Given the relatively small
20\arcsec $\times$20\arcsec\ (0.89~kpc$^2$) region containing all of
this activity, the star formation rate per unit area is at least
$\simeq$ 2 \mo\ yr$^{-1}$ kpc$^{-2}$. The {\it Spitzer Space
Telescope} 24 $\mu$m flux of 4.9~Jy \citep{engelbracht05} implies a
much higher star formation rate of 6.8 M$_\odot$~yr$^{-1}$ using the
calibration of \citet{calzetti2007}, indicating that much of the star
formation is heavily obscured \citep{vacca2002}.

He~2-10 contains an \ion{H}{1} mass of $4.9\times10^8$ \mo\
\citep{kobulnicky1995} and molecular mass of $2.9\times10^8$ \mo\
yielding a helium-corrected gas mass fraction of $\leq$0.20,
consistent with Henize~2-10 being a transition object between gas-rich
dwarfs and larger, more gas-poor spiral galaxies.
Beam-averaged \ion{H}{1} column densities in the central region vary
from few $\times10^{20}$~cm$^{-2}$ to as high as 2$\times10^{22}$
cm$^{-2}$ \citep{baas1994, kobulnicky1995, meier2001, santangelo2009}
corresponding to extinctions up to $A_V$=20 mag for typical Galactic
gas-to-dust ratios \citep{bohlin1978}.  This is consistent with the
localized high extinctions of $A_V$=10--30 mag inferred from mid-IR
silicate absorption and IR Brackett emission-line methods
\citep{phillips1984, kawara1989, beck1997} but much higher than the
$A_V$=1--2 mag derived toward optically visible star clusters and
\ion{H}{2} regions \citep{vacca1992}.  These disparate results warn
that the nuclear conditions are complex on small angular scales with
large variations along the line of sight.  This high dust and
molecular content is consistent with the strong CO emission and with
chemical abundance measurements showing a metallicity near the solar
value \citep{kkp99}.  \citet{darling2008} detect H$_2$O ``kilomasers''
in the nucleus, indicating the presence of many massive star-forming
regions.  Single-dish radio continuum measurements over a range of
frequencies show a dominant non-thermal component, indicating the
presence of significant supernova activity \citep{allen1976}.
H$\alpha$ images of Henize~2-10 show multiple ionized shells and
filaments typical of actively starforming dwarf galaxies
\citep{marlowe1995}.  High-resolution H$\alpha$ spectroscopy reveals
multiple kinematic components, with the fastest reaching $\Delta V=$250 \kms\
\citep{mendez1999}.

Blueshifted interstellar absorption lines seen against the nucleus
indicate an outflow of low-ionization material along the line of sight
\citep{schwartz2006, johnson2000}.  Published outflow speeds vary by a
factor of two owing to uncertainties regarding the systemic velocity,
low spectral resolution, and the wavelength ambiguity caused by the
uncertain positioning of the continuum source within a large spectral
aperture. Optical echelle spectroscopy obtained with HIRES at the
W. M. Keck Observatory offer the most accurate measurements to date
\citep{schwartz2005}.  Excited stellar \ion{Mg}{1} triplet
$\lambda\lambda\lambda$ 5167.32, 5172.68, 5183.60 \AA\ absorption
yields a heliocentric systemic velocity of $869 \pm 3$\kms, consistent
with the 21-cm and CO data.  The narrow slit and seven \kms\ resolution reveal
four blueshifted interstellar \ion{Na}{1} $\lambda\lambda$ 5889.95,
5895.92 \AA\ components in addition to a stellar absorption component
at the systemic velocity. The fastest (and weakest) of these reaches a
maximum velocity of -180 \kms. The average outflow speeds of the four
components are $-34\pm2$, $-65\pm2$, $-93\pm4$, and $-128\pm2$ \kms.

The maximum \ion{H}{1} rotation velocity observed by
\citet{kobulnicky1995} provides a lower limit on the dynamical mass of
He~2-10. For the adopted distance of 9.8~Mpc and an inclination of 30
degrees, the dynamical mass within 2.1~kpc is $M_{dyn}= 1.18 \times
10^{10}~M_{\odot}$.  The modified isothermal sphere model of
\citet{ferrara2000} for the mass distribution implies a halo scale
radius of 4.8~kpc and a central density of $9.5 \times
10^6~M_{\odot}~kpc^{3}$. We estimate the escape velocity at the disk
radius of 2.1~kpc is at least 150~km~s$^{-1}$ for a disk inclination
of 30 degrees. Hence little of the observed outflow exceeds the escape
velocity. A decrease in the covering fraction of the low-ionization
gas with radius \citep{martin2009} and/or increasing ionization
fraction with radius, however, might render higher velocity gas
difficult to detect.

\section{Data Acquisition and Reduction}
\subsection{{\it Chandra} Observations}

The {\it Chandra X-ray Observatory} observed Henize~2-10 with the \cxo\ CCD
Imaging Spectrometer (ACIS) on 2001 March 23 for 20.02 ksec.  The
galaxy was placed on the back-illuminated chip ACIS-S3 (also known as
chip~7).  These data, as well as the fields on the ACIS-I and ACIS-S5
chips, are available in the archive under sequence number 600209.

The data were processed by the \cxo\ X-ray Center (CXC) using version
R4CU5UPD14.1 of their software. CIAO, version 2.3, with CALDB, version
2.4, was used for the majority of the data processing unless otherwise
noted.  We checked the positional accuracy of the chip~7 \cxo\ sources
using our R-band image.  For the 9 X-ray sources with optical
detections, the rms positional discrepancy was found to be
0\farcs5. The light curve was inspected, and no periods of high
background occurred during the observation. No additional
re-processing was required.

Spectra were extracted using pulse invariant data values to account
for gain variations between nodes.  For each source region, we
weighted the appropriate CXC spectral response files by the
distribution of their areas within the aperture.\footnote{We used the
calcrmf/calcarf software package contributed by Jonathon McDowell
which is available from the CXC website --
http://asc.harvard.edu/cgi-gen/cont-soft/soft-list.cgi.}  The maximum
difference between our area-weighted response files and single
response files extracted at the flux-weighted centroid of the aperture
was a few percent. Weighting the response files by the distribution of
counts within the aperture produced results indistinguishable from the
area-weighted responses.

Unbinned images were extracted from the time-filtered events file in
four carefully chosen bands: soft (S) 0.3--0.7~keV, medium (M) 
0.7--1.1~keV, hard (H) 1.1--6~keV, and Total (T) 0.3--6~keV.
Emission in the soft band is somewhat attenuated because the Galactic
foreground column, $N_{H} = 4.89 \times 10^{20}\col$ (Hartmann \&
Burton 1996), corresponds to optical depth unity at 0.35~keV.  The galaxy
was not detected at energies above 6~keV, the energy adopted for the
hard band cutoff. Background images were extracted from the background
events file in these bands.

Each of the four X-ray images was smoothed using the adaptive
smoothing algorithm of H. Ebeling \& V. Rangarajan as implemented as
CSMOOTH in CIAO.  The smoothing scales are automatically adjusted
to achieve a minimum S/N ratio of 1.8 and a maximum S/N ratio of 2.2
per pixel.  Strong point sources are effectively unchanged by the
smoothing process, while the contrast of weak, diffuse emission is
enhanced.  Each of the four background images was smoothed (using 
the smoothing map generated by CSMOOTH from the on-source images) and 
subtracted from the source images.

Since adaptive smoothing does not preserve photon statistics in a
straightforward manner, we used the adaptively smoothed images only to
produce images for presentation and to define large apertures for the
extraction of diffuse component spectra.  All quantitative analyses
were performed by extracting photons from the events file or
unsmoothed images and by estimating their significance using Poisson
statistics.

\subsection{Point Source Identification}

\begin{figure}
\plotone{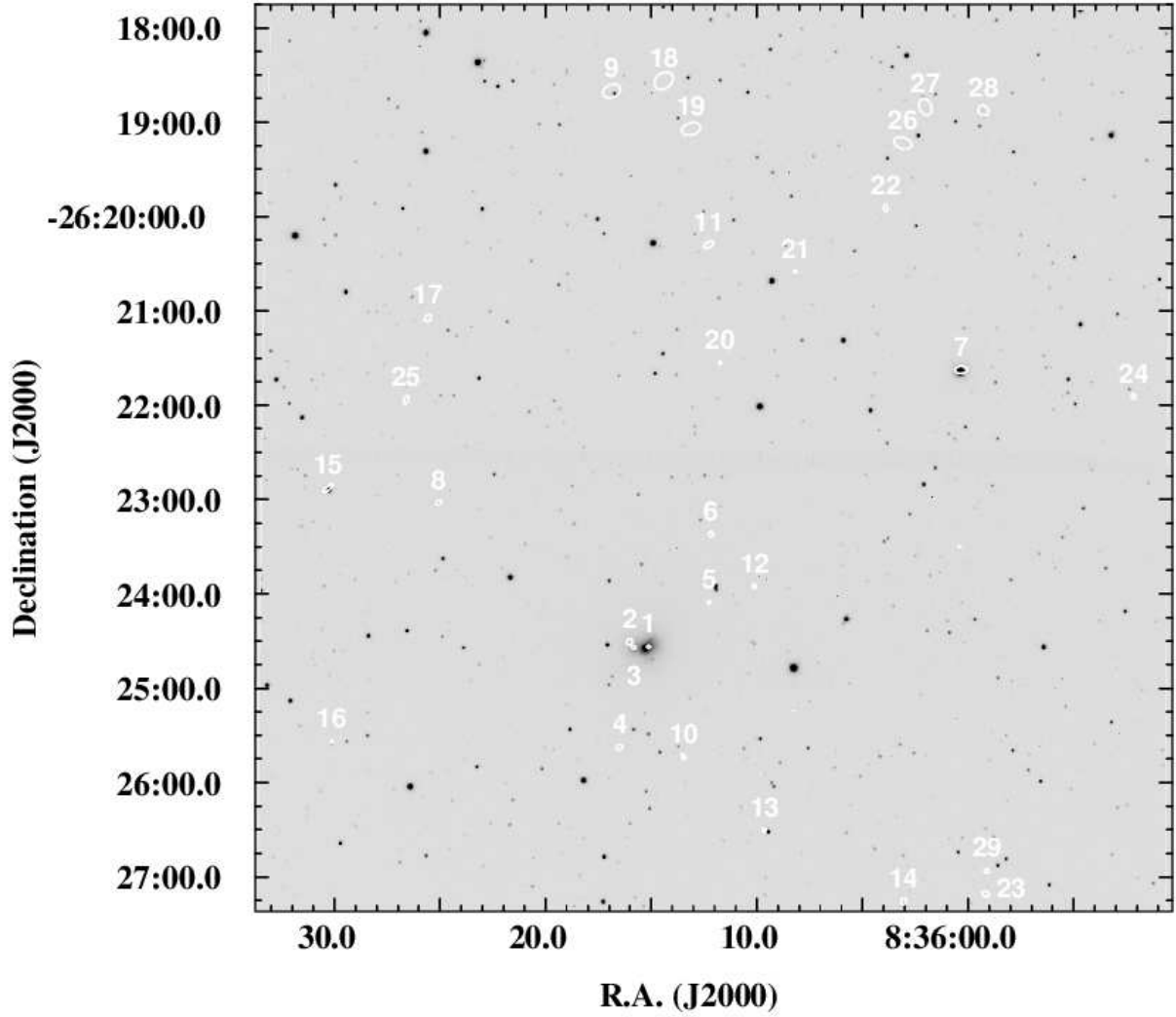}
\caption{Grayscale optical 6550 \AA\ image with
compact X-ray sources ({\it numbered white ellipses}) detected in the 0.3 -- 6 keV band
near Henize~2-10.
Object labels correspond to the X-ray sources listed in Table~\ref{src.tab}.  }
\label{X-Rband} 
\end{figure}

\begin{figure}
\plotone{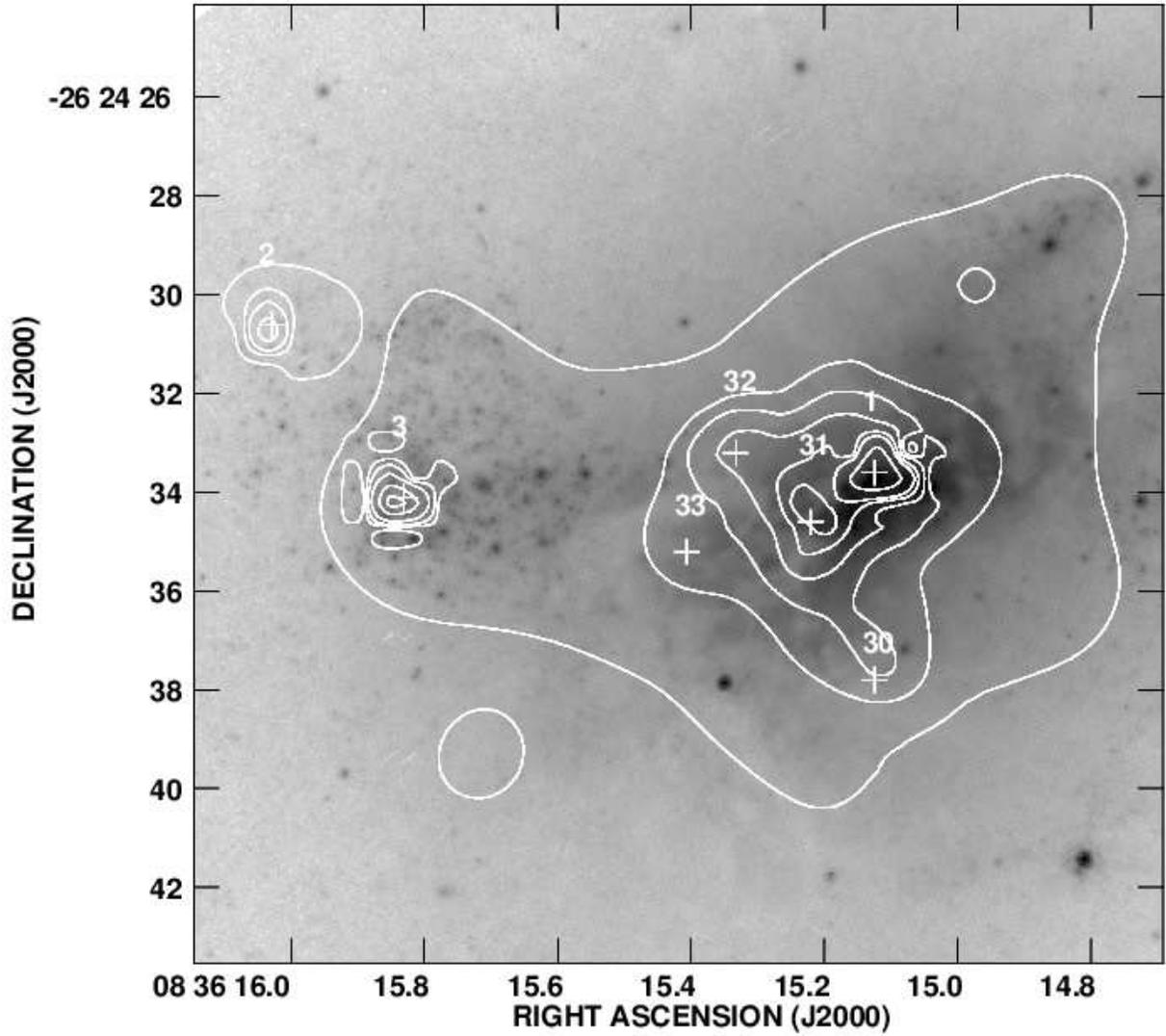}
\caption{X-ray point sources ({\it white crosses}) near the nucleus of Henize~2-10.  
The grayscale shows a logarithmic
representation of the Hubble Space Telescope F814W image 
from \citet{johnson2000}, and the contours show the
adaptively-smoothed \cxo\ ACIS total band 0.3--6 keV image.  Contours
levels are drawn every factor of 2 at 0.42, 0.85, 1.7, 3.5, 7.0, 14.0, 28.0 
photons~s$^{-1}$ arcmin$^{-2}$.  }
\label{X-HST} 
\end{figure}

\begin{figure}
\plotone{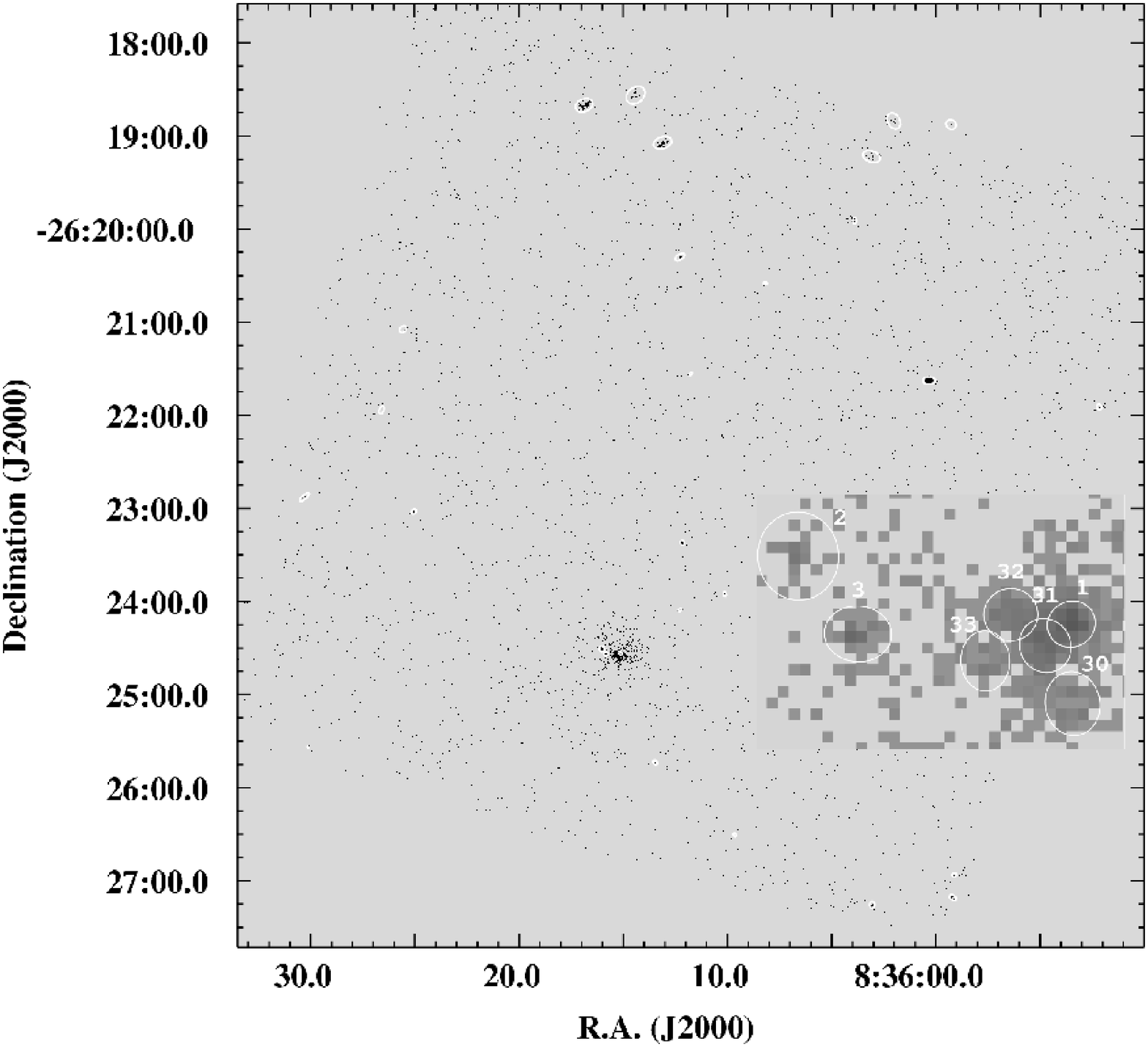}
\caption{An unsmoothed total band 0.3 -- 6 keV X-ray image 
the ACIS S-3 chip with He~2-10 near the center. White ellipses are the
same as in Figure~\ref{X-Rband}.  The inset shows a magnified view
near the nucleus of He~2-10 with objects 1,2,3 and 30--33 numbered.       }
\label{Xraw} 
\end{figure}

The CIAO 4.1 detection algorithms CELLDETECT and WAVDETECT were used
to identify compact X-ray sources on the ACIS S3 chip over the energy
range 0.3 -- 6 keV.  We used WAVDETECT with wavelet scales between 1
and 32 pixels as the as the primary detection tool, and this yielded
29 probable X-ray sources, the weakest of which had a significance
level of 1.7$\sigma$.  Three sources (\#'s 1,2,3) lie projected
against the stellar component of He~2-10, while the remainder may be
associated with field stars or background galaxies.  CELLDETECT found
a subset of these 29 sources, but also yielded four additional sources
not identified by WAVDETECT (\#'s 30 -- 33).  These four lie within
10\arcsec\ of the nucleus where the diffuse X-ray emission is strong,
have sizes exceeding the nominal PSF size, and, we show later, 
are likely to be localized peaks in the diffuse thermal
emission.  Table~1 lists the positions of the 29+4 compact sources
detected.  Figure~\ref{X-Rband} shows these sources numbered 1--29
overlaid on an optical 6550 \AA\ image.  Sources 1,2,3 and 30--33 are
better seen in Figure~\ref{X-HST} which shows an adaptively smoothed
X-ray depiction ({\it contours}) of the Henize~2-10 nucleus overlaid on
a 400~s F814W image ({\it grayscale}) from the Hubble Space Telescope
archive \citep{johnson2000}.  Figure~\ref{Xraw} shows an unsmoothed
0.3 -- 6 keV X-ray image of the ACIS S3 chip with the elliptical source
regions, as in Figure~\ref{X-Rband}.  The inset in this Figure shows a
zoomed view near the nucleus that includes sources 1,2,3 and 30--33.

We measured counts in each of the four energy bands (Total, Soft,
Medium, and Hard) using elliptical apertures containing probable
source pixels as identified by WAVDETECT.  Annuli surrounding each
aperture were defined manually and used for measuring and subtracting
the background.  Table~1 lists the net counts detected in each band
along with the ratio of the source size to the nominal PSF.  Sources
1,2,3, 10 and 30--33 have PSF ratios near 2 or larger, suggesting that
they may be either extended sources or blends of multiple point
sources.

Our detection threshold of $\sim$4 counts in 20 ksec in the Total
0.3--6 keV band on the {\it Chandra} $ACIS$ corresponds to an X-ray
flux of 1.5$\times10^{-15}$ based on the WebPIMMS mission count rate
simulator. At or above this flux level we would expect an X-ray source
density of $\sim$0.8 sources per square arcminute, based on the
results of \citet{mushotzky00}, \citet{moretti03}, and \citet{kim07}.
Within the $\sim$0.25 square degree region encompassing He~2-10, we
would therefore expect, on average, 0.2 background X-ray sources.
Statistically, the three (possibly seven) compact sources projected
onto He~2-10 are likely to lie within the galaxy itself.

\subsection{Optical imaging}

Optical images of He~2-10 were obtained the night of 2001 March 28 with
the 3.5~m WIYN\footnote{The WIYN Observatory is a joint facility of
the University of Wisconsin-Madison, Indiana University, Yale
University, and the National Optical Astronomy Observatory.}
telescope and the MiniMosaic detector.  Seeing was highly variable,
and non-photometric conditions prevailed.  Exposures of 60 s were
obtained using a filter of width 385 \AA\ centered at 6550 \AA.  Two
exposures with the smallest point spread functions were combined to
produce an image with a mean stellar FWHM of 1.2\arcsec. 
A rough R-band flux calibration was performed on the image
using the magnitudes of 8 stars in the field based on the Guide Star
Catalog F-band magnitudes.  The typical uncertainty is 0.5 mag.  Poor
weather and low target elevation prevented $H\alpha$ narrow band
exposures from being taken.  We instead obtained a
continuum-subtracted $H\alpha$ image of Henize~2-10 which appeared in
\citet{mendez1999} (kindly given to us by C. Esteban.)  The
$H\alpha$ image is flux calibrated using the calibration of \citet{mendez1999}.

\subsection{Radio Continuum Observations}

Radio continuum observations of Henize~2-10 at 20 cm (1420 MHz) and 2
cm (15 GHz) were obtained with the Very Large Array in the B
configuration and published previously in \citet{kobulnicky1995} and
\citet{kj99}.  Full details may be found in those
papers.  We used the 20-cm data mapped with a synthesized beam of
3\arcsec\ FWHM to search for radio sources in the field surrounding 
He~2-10 that correspond to the detected X-ray and optical sources.
Other than Henize~2-10 itself, there are no 20-cm radio sources
within the 8\arcmin\ \cxo\ field to a 3$\sigma$ limit of 0.6 mJy.

\section{X-ray Properties}
\subsection{Nature of the Point Sources}

Sources 1--3 and 30--33 appear superimposed on diffuse emission near the
nucleus of Henize~2-10, and all have PSF ratios between 2 and 4,
while the remainder of the 26 compact sources further from the
nucleus have PSF ratios not far from unity.  These nuclear objects
may either be genuine point sources (i.e., X-ray binaries or
collections of X-ray binaries) projected onto regions of diffuse gas,
or they may be local maxima in the diffuse emission. The second
possibility better explains their larger sizes compared to the field
sources.  Their X-ray colors help to discriminate between these
possibilities.

\begin{figure}
\plotone{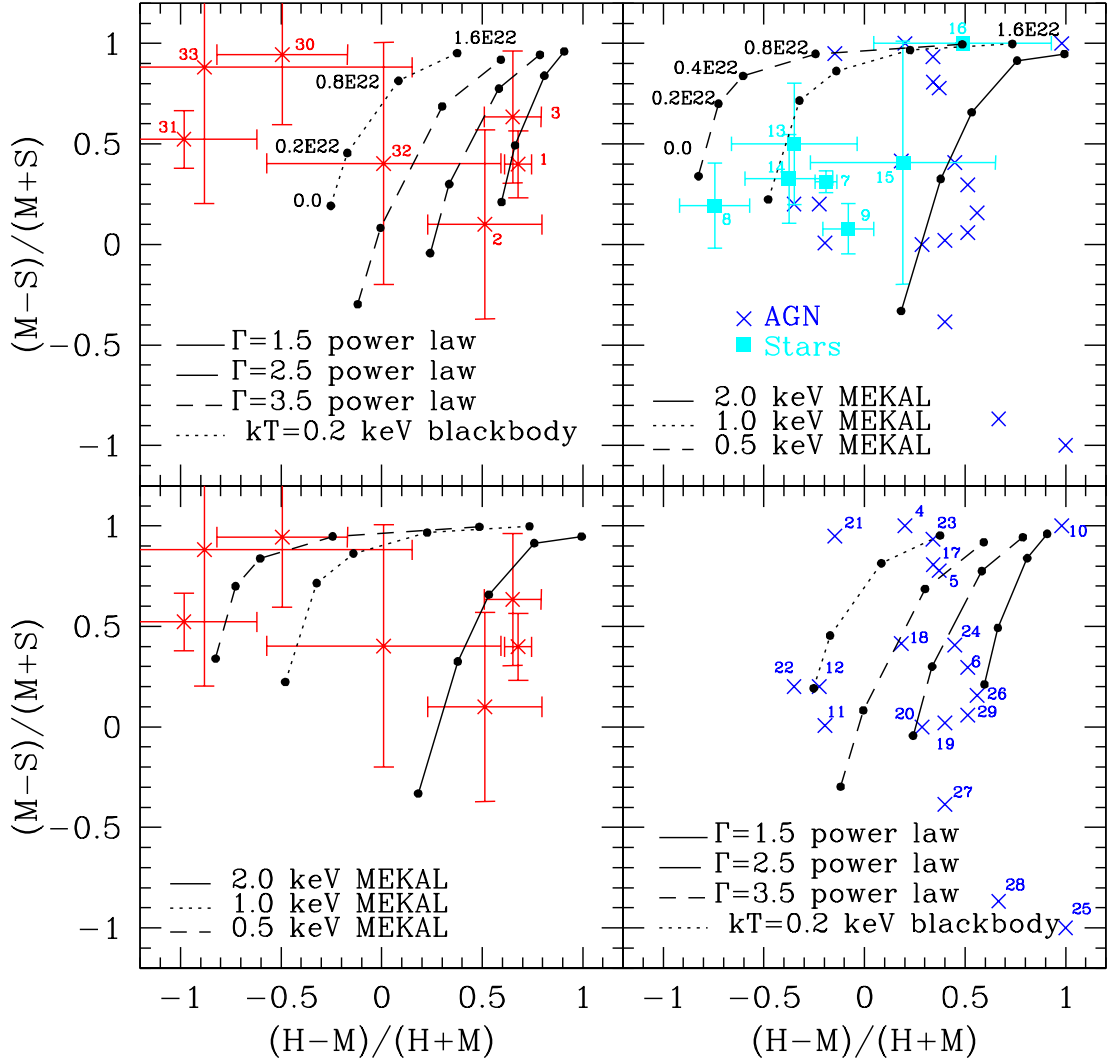}
\caption{Two-color diagram for X-ray point sources.
        Lines represent the colors of model spectra.  
        Upper left: The seven sources projected against the nucleus of Henize~2-10. 
        Power laws with photon indices of $\Gamma$=1.5, $\Gamma$=2.5, 
        $\Gamma$=3.5, and a blackbody spectrum with a temperature kT=0.2 keV 
        are shown for several foreground  \ion{H}{1} column densities of 0.0, 0.2$\times10^{22}$, 
	 0.8$\times10^{22}$,  and 1.6$\times10^{22}$ cm$^{-2}$.
	Lower left: The seven compact sources.  MEKAL thermal plasma models with temperatures kT=0.5, kT=1.0 and kT=2.0 keV 
        for several foreground  \ion{H}{1} column densities are shown. 
       Upper right and lower right: Other X-ray sources which lie outside the disk of He~2-10.  Thermal plasma models 
	and power law models for
	a range of column densities are shown. 
        Filled squares denotes objects identified as probable stars in Table~\ref{src.tab} 
        on the basis
        of X-ray to optical fluxes, $\log(f_X/f_V)<-0.5$, while crosses denote 
        probable AGN with $\log(f_X/f_V)>-0.5$. Error bars are omitted on AGN for clarity.    }
\label{2color} 
\end{figure}

Figure~\ref{2color} shows an X-ray two-color diagram formed from
ratios of count rates in the hard, medium, and soft bandpasses:
$(M-S)/(M+S)$ and $(H-M)/(H+M)$.  The left panels include only the
seven sources projected against the nucleus, while right panels show
the other 26 sources.  Lines and filled circles illustrate the X-ray
colors for various power law, thermal plasma (MEKAL), and blackbody
spectral energy distributions for four different foreground absorbing
columns: 0.0, 0.2$\times10^{22}$, 0.8$\times10^{22}$, and
1.6$\times10^{22}$ cm$^{-2}$. Sources 1--3 are consistent
with $\Gamma=1.5$ power law spectra behind modest absorbing columns.  These three
objects are strongest in the hard band, consistent with heavily
absorbed spectra.  Sources 2 and 3 are too faint for meaningful
spectral fits, but lower limits on their luminosities can be made
assuming a {\it minimum} column density of $N_H=0.05\times10^{22}$
\col\ for the Galactic foreground and a power law spectrum with photon
index $\Gamma=1$ where the number of photons per energy bin is given
by N(E)$\propto E^{-\Gamma}$.  For these values, sources 2 and 3
have probable unabsorbed 0.3--8 keV luminosities of
$>$3.4$\times10^{38}$ erg s$^{-1}$ and $>$1.7$\times10^{38}$ erg s$^{-1}$
respectively at 9.8 Mpc.  Although not quite as luminous as the
``superluminous'' X-ray sources found in nearby galaxies ($>10^{39}$
erg s$^{-1}$; e.g., \citet{roberts2004}), these objects are likely to
be powered by accretion onto stellar mass black holes.  However, the
uncertainties in the foreground column and the photon energy
distribution would allow for intrinsic luminosities several times
greater than these estimates if $N_H$ and $\Gamma$ were larger than
the values adopted here.  We note that both of these objects 
appear to lie within the older starburst region ``B'' \citep{vacca1992} 
several arcseconds to the east of the dominant starburst core region (``A''). 

By contrast with sources 2 \& 3, sources 30--33 are
strongest in the medium band, consistent with the color expected of soft
thermal plasma.  The upper right panel of Figure~\ref{2color} plots
the colors of sources 4--29.  We tentatively identify these sources as
either stars or active galactic nuclei (AGN) based on the ratio of
their X-ray to optical fluxes, following \citet{krautter1999}.  The
ratio of X-ray to optical fluxes is $f_X/f_V$, where
$f_X$(erg~s$^{-1}$ cm$^{-2}$) = 3.6$\times10^{-16}$ $\times C_{tot}$
where $C_{tot}$ is the number of detected 0.3--6.0 keV X-ray photons
in 20 ks, and $f_V$(erg~s$^{-1}$ cm$^{-2}$)=1.7$\times10^{-6} \times 2.5^{-R}$
where R is the R-band magnitude.  Table~\ref{src.tab} lists objects
with $\log(f_X/f_V) < -0.5$ as probable stars and objects with
$\log(f_X/f_V) > -0.5$ as probable AGN.  In Figure~\ref{2color},
probable stars appear as filled squares and probable AGN appear as
crosses.  Solid dots connected by lines indicate the colors of
MEKAL thermal plasmas with temperatures of 0.5, 1.0, and 2.0 keV
attenuated by absorbing columns of 0.0, 0.2$\times10^{22}$,
0.4$\times10^{22}$, 0.8$\times10^{22}$, and 1.6$\times10^{22}$
cm$^{-2}$.  The candidate stars occupy a region of relatively low
column density, as might be expected for Galactic foreground objects.
The candidate AGN are distributed throughout the diagram, consistent
with a range of power law slopes and absorbing columns. The
strongest source in the field is object 7, the star
GSC~06578-03010.  It is a previously known X-ray source,
RX~J0836.0-2621.

\subsection{Diffuse versus Point Source Flux Contributions}

The compact sources 1--3 represent only a small fraction of the X-ray
flux of Henize~2-10.  We performed photometry on the X-ray images
using an 80\arcsec\ diameter aperture centered on the nucleus to
obtain a total flux for the galaxy.  The size of this aperture was
chosen to include all of the diffuse emission seen in the smoothed,
background-subtracted total band (0.3--6.0 keV) image.  This aperture
includes the seven compact sources in the main body of Henize~2-10 and
none of the other 26 sources in the field.  The background flux was
estimated using an annular region extending 15\arcsec\ beyond the
circular aperture.  Table~\ref{phot.tab} lists the results of this
photometry in each band.  Table~\ref{phot.tab} also indicates the
fraction of diffuse emission after subtraction of the flux from the
compact sources 1--3.  We find that 84$\pm$4\% of the emission in the
total 0.3--6.0 keV band comes from diffuse emission.  In the hard
1.1--6.0 keV band, 69$\pm$9\% of the emission is diffuse.  In the soft
and medium bands, the fraction of flux from the diffuse X-ray medium
is over 90\%.  Thus, the X-ray flux of Henize~2-10 appears to be
dominated by the diffuse plasma component even in the highest energy
band.

\subsection{X-ray Spectrum of the Nuclear Source: An Intermediate-Mass Black Hole?}

\begin{figure}
\plotone{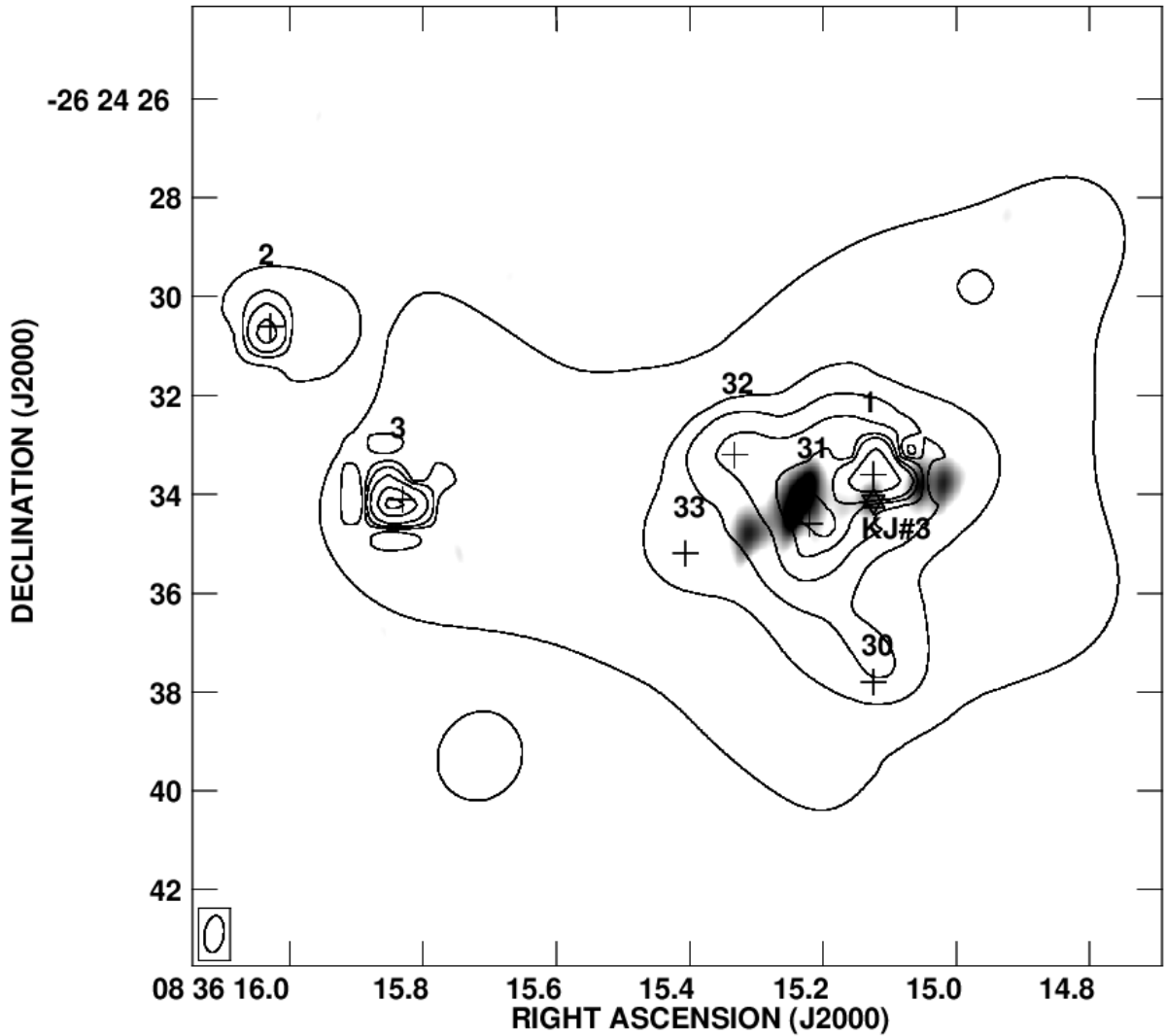}
\caption{X-ray  \cxo\ ACIS total band 0.3--6 keV image of the nucleus of Henize~2-10
with contours as in Figure~\ref{X-HST}.
The grayscale shows a logarithmic
representation of the 2~cm Very Large Array radio continuum maps
from \citet{kj99}.  A star marks the position of
radio source KJ3 from \citet{kj99} which lies 0.25\arcsec\ south, but within
the positional uncertainties of the dominant X-ray point source \#1. 
 }
\label{6-ALL} 
\end{figure}

\citet{jk2003} noted the non-thermal nature of the spectral energy
distribution of the pointlike radio source in the nucleus (source \#3
of \citet{kj99}) but did not discuss the nature of this source.
Figure~\ref{6-ALL} shows the X-ray contour image from
Figure~\ref{X-HST} overlaid on a grayscale representation of the 2 cm
(15 GHz) radio map from \citet{kj99}.  Radio source KJ3, marked by a
star, lies 0.25\arcsec\ south of the brightest nuclear X-ray source,
\#1.  Within the uncertainties of the radio and X-ray data,
the two sources are coincident.  The optical and mid-IR images of
\citet{cabanac2005} show that there is no bright optical or infrared
source at this location.\footnote{\citet{cabanac2005} identify and
correct erroneous astrometry in the $HST$ images that were used to
compare optical and radio/IR positions in \citet{kj99},
\citet{vacca2002}, \citet{jk2003}, and in some earlier works. The net
result is that the $HST$ images should be shifted 1{\arcsec}.2 {\it
west} relative to the radio and IR images in Figures from those prior
publications. }

\begin{figure}
\plotfiddle{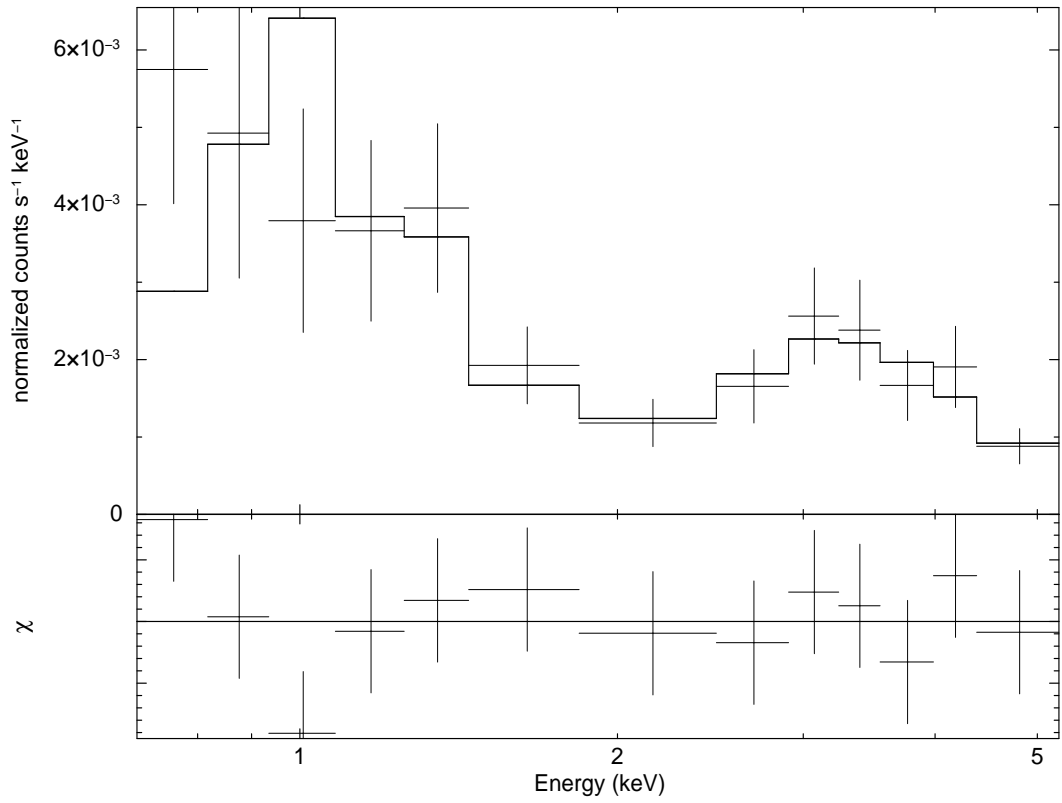}{3.0in}{-90}{300}{400}{0}{0} 
\caption{X-ray spectrum of the nuclear point source \#1
and best-fit absorbed power law+MEKAL model and residuals (Model 
A in Table~\ref{spectra_nuc.tab}). Error bars represent 1-sigma uncertainties. 
The fit shown is for power law photon index $\Gamma=4$,
plasma temperature $kT=0.4$ keV and foreground absorbing column
density of $9.7\times10^{22}$ cm$^{-2}$.
   }
\label{src1.spec} 
\end{figure}

We extracted an X-ray spectrum of the region within 1\arcsec\ ($\sim2$
pix) of the dominant X-ray source, \#1, using a neighboring annular
region to estimate the local X-ray background.  Other choices of local
background yield similar results.  The resulting spectrum, shown in
Figure~\ref{src1.spec}, was then analyzed in Xspec (version 11.3). No
single-component spectral models produce acceptable fits to the data.
Reasonable models require at least two spectral components to
reproduce both the low-energy peak near 1 keV and the secondary peak
near 3 keV.  This secondary high-energy peak is most pronounced in
small 2\arcsec\ diameter spectral apertures centered on the nucleus
and becomes weaker relative to the rest of the spectrum as the
aperture size is increased.  This indicates that the high energy
photons near 3 keV are concentrated near the nucleus, comprising a real
spectral feature of the nuclear source.  The observed 0.3--6.0 keV
flux from the nuclear source is 7.6$\times10^{-14}$ erg s$^{-1}$
cm$^{-2}$.  Given the high optical extinction toward the nucleus of
Henize~2-10 ($A_V$=10--30 mag; \citet{phillips1984, kawara1989})
X-rays are likely to provide the best estimate of the bolometric
luminosity of the central source.

We performed spectral fits to the nuclear source assuming a minimum
foreground Galactic column of 0.05$\times10^{22}$ cm$^{-2}$ of solar
abundance (component WABS\footnote{Adopting a model for the absorbing
ISM based on the more recent X-ray cross sections of \citet{wilms00}
(TABS in Xspec) changes the best fit parameters negligibly.} in
Xspec).  We fit the X-ray SED with spectral models that were linear
combinations of a solar-abundance thermal plasma component (MEKAL with
variable foreground, $N_H$, and variable temperature, $kT$)
superimposed on either a power law component (POW with variable $N_H$
and photon index, $\Gamma$) or a second thermal component (variable
$N_H$ and variable $kT$).  In all cases, the second component required
very high foreground columns of more than $2\times10^{22}$ cm$^{-2}$
in order that it contribute at 3 keV but not at lower energies.  Such
high column densities are consistent with the beam-averaged molecular
gas column densities toward the nucleus \citep{kobulnicky1995} and
correspond to optical extinctions of $A_V$=20 for typical Galactic
ratios $N_{H_2}/A_V=9.4\times10^{20}$ cm$^{-2}$ mag$^{-1}$
\citep{bohlin1978}.
 
The best fitting models have reduced $\chi^2 \simeq 1$ and involve an
absorbed MEKAL thermal plasma component and a heavily absorbed power
law component (WABS(MEKAL + WABS*POW)).  Table~\ref{spectra_nuc.tab}
summarizes the parameters, $\chi^2$ values, and implied unabsorbed
luminosities for a range of power law photon index, $\Gamma$, ranging
from 1 to 4.  In these fits the temperature of the thermal component
is fixed at 0.4 keV, while the normalizations and absorbing columns of
each component are allowed to vary.  Model A which has a steep power
law index of $\Gamma=4$ and very high absorbing column of
24$\times10^{22}$ cm$^{-2}$ produces the best fit, having reduced
$\chi^2$=0.89 for 9 degrees of freedom.  Such a steep power law index
is consistent with the class of ``super soft'' extragalactic X-ray
sources \citep{cagnoni2003} and some of the ultraluminous X-ray
sources in the compilations of \citet{winter2006} and \citet{berghea2008}.
Models B and C with $\Gamma=2$ and $\Gamma=1$, respectively, also
yield acceptable fits with $\chi^2$=1.10 and $\chi^2$=1.33.
Substituting a 500--700 eV blackbody component for the power law
component produces similar acceptable fits.  In general, steeper power
law photon indices require larger absorbing column densities. In this
series of models, the normalization of the power law component ranges
from a factor of 10 to a factor of two larger than the MEKAL
component, indicating that the power law component dominates the SED.
Neither the slope of the photon index, $\Gamma$, nor the foreground
column density are well constrained given that there are only 142
total counts in the X-ray spectrum.  However, the best fit parameters
represent reasonable physical conditions similar to those observed in
studies of X-ray binaries \citep{remillard2006} and extragalactic
X-ray point source populations \citep{strickland2000, zezas2002}.

Any of the the models in Table~\ref{spectra_nuc.tab} imply an {\it
unabsorbed} 0.3--6.0 keV X-ray luminosity in the central 2\arcsec\ of
$L_{X(0.3-6.0)}\simeq10\times10^{38}$ erg~s$^{-1}$ to as much as
$L_{X(0.3-6.0)}\simeq1200\times10^{38}$ erg~s$^{-1}$ for a distance of
9.8 Mpc.  Such luminosities are similar to the ultraluminous (ULX)
X-ray point sources with $L_{X}=10^{39}$ -- $10^{41}$ erg~s$^{-1}$ which
are common in nearby galaxies \citep{colbert2002, swartz2004,
winter2006, berghea2008}.  Some galaxies like the Antennae
(NGC~4038/39) harbor several \citep{zezas2002}.

The physical nature of ULX sources has not been firmly established.
They may be stellar mass black holes accreting at super-Eddington rates
or radiating non-isotropically (i.e., beamed) \citep{king2001,
kording2002, rappaport2005, king2009}, or they may be intermediate-mass black
holes with $M$=20--10000 M$_\odot$ \citep{colbert1999, miller2004}.
Some combination of these possibilities is likely.  ULX sources are
usually associated with regions of star formation, often have optical
counterparts \citep{ramsey2006}, and are uncommon \citep{irwin2004} but
not non-existent \citep{ptak2006} in early-type galaxies.

If we adopt an X-ray luminosity of ${\sim}1\times10^{40}$ erg~s$^{-1}$
for the isotropic luminosity of an accreting object in Henize~2-10,
then the implied mass of the central object is $M\geq100$ \mo, where
the inequality covers the possibility that the object may accrete at
rates below the Eddington rate. The formulation of \citet{merloni2002}
expressing the black hole mass in terms of the X-ray and radio
luminosity gives 46 $M_\odot$.  This mass is much less than the
predicted mass of a nuclear black hole, $M=few \times10^5$ \mo,
obtained by using the \ion{H}{1} velocity dispersion of $\sigma_v=46$
\kms\ \citep{kobulnicky1995} with the well-established black hole
mass-velocity dispersion correlation \citep{gebhardt2000,
tremaine2002}.  If the nuclear X-ray source marks the presence of a
black hole near the center of Henize~2-10, either the 
accretion rate is well below the Eddington rate, or the canonical
$M_{BH}-\sigma$ relation may not obtain for such a low-mass galaxy.

\begin{figure}
\plotone{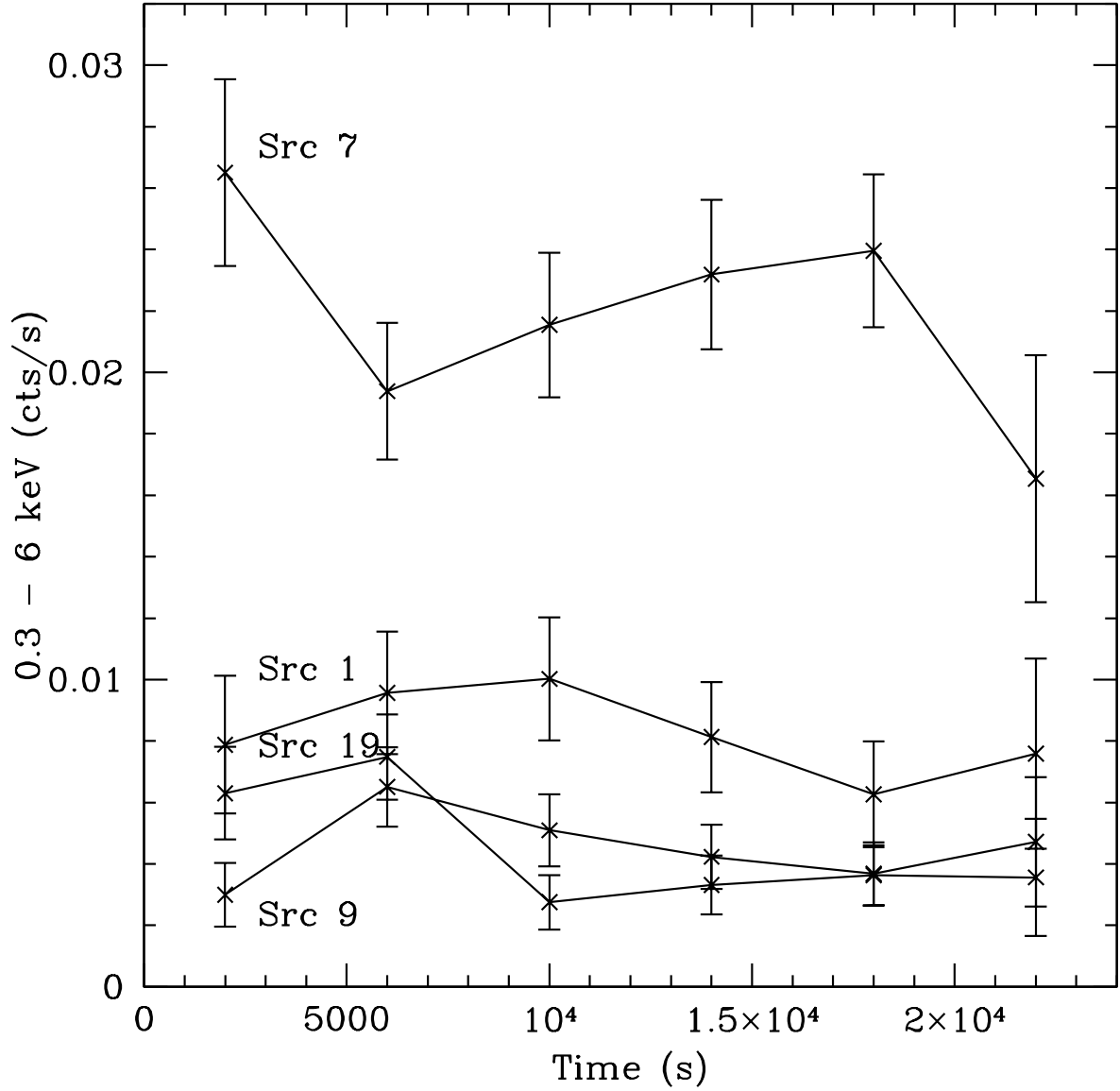}
\caption{X-ray 0.3--6.0 keV light curve of the nuclear point source \#1 and 
three other field X-ray sources from Table~1.   }
\label{lc} 
\end{figure}

We extracted time-binned count rates for the central source over the
full 0.3--6 keV energy range using bin widths of 2000 s and 4000 s to
search for variability.  Figure~\ref{lc} shows the light curve of the
nuclear source \#1 compared to three other field X-ray sources (\#7,
\#9, \#19) with sufficiently high count rates to construct a 
comparison light curve.  Plotted uncertainties reflect photon
statistics in the source and background regions.  None of the sources
show variability at greater than the 1--2$\sigma$ level over the 20
ksec period of observation.  The nuclear source varies by 40\%
(2$\sigma$), making the evidence for variability marginal.

Another indication of the source nature is the emission at radio
wavelengths.  \citet{miller2005} and \citet{lang2007} report Very
Large Array centimeter-wave radio studies of ULX sources in
Holmberg~II and NGC~5408, respectively, and show that the radio
emission is extended over scales of 35--60~pc and has non-thermal
spectral indices $\alpha$=-0.5 -- -0.8 ($S_\nu \propto
\nu^\alpha$) characteristic of optically thin synchrotron emission in
both cases.  Such extended emission implies a large ionized nebula.
Radio source \#3 in He~2-10 \citep{jk2003} is also non-thermal ($\alpha
\sim -0.9\pm0.5$), and, while the angular size is unresolved in the
$\sim$1\arcsec\ data, implying a diameter smaller than 50 pc, the
non-detection of this source with few milliarcsecond resolutions rules
out a compact supernova or supernova remnant and suggests an extended
morphology \citep{ulvestad2007b}.  The ratio of radio to X-ray fluxes
for this source, $R_X$ \citep{terashima2003}, is

\begin{equation}
R_X = {{\nu L_\nu(5~GHz)} \over {L_X(2 - 10~keV)} } = 
{{2.9 \times 10^{-16} erg~s^{-1}~cm^{-2}} \over{1.7 \times 10^{-13} erg~s^{-1}~cm^{-2}  } }
  = 1.7\times10^{-3} .  
\end{equation}

\noindent This is in the range of other ULX sources and low-luminosity AGN
tabulated by \citet{neff2003} ($R_X=1\times10^{-2} - 2\times10^{-3}$),
but higher than X-ray transients ($R_X<10^{-5}$) and lower than a
supernova remnant like Cas~A ($R_X=2\times10^{-2}$).

\subsection{Extended X-ray Emission}

\begin{figure}
\plotone{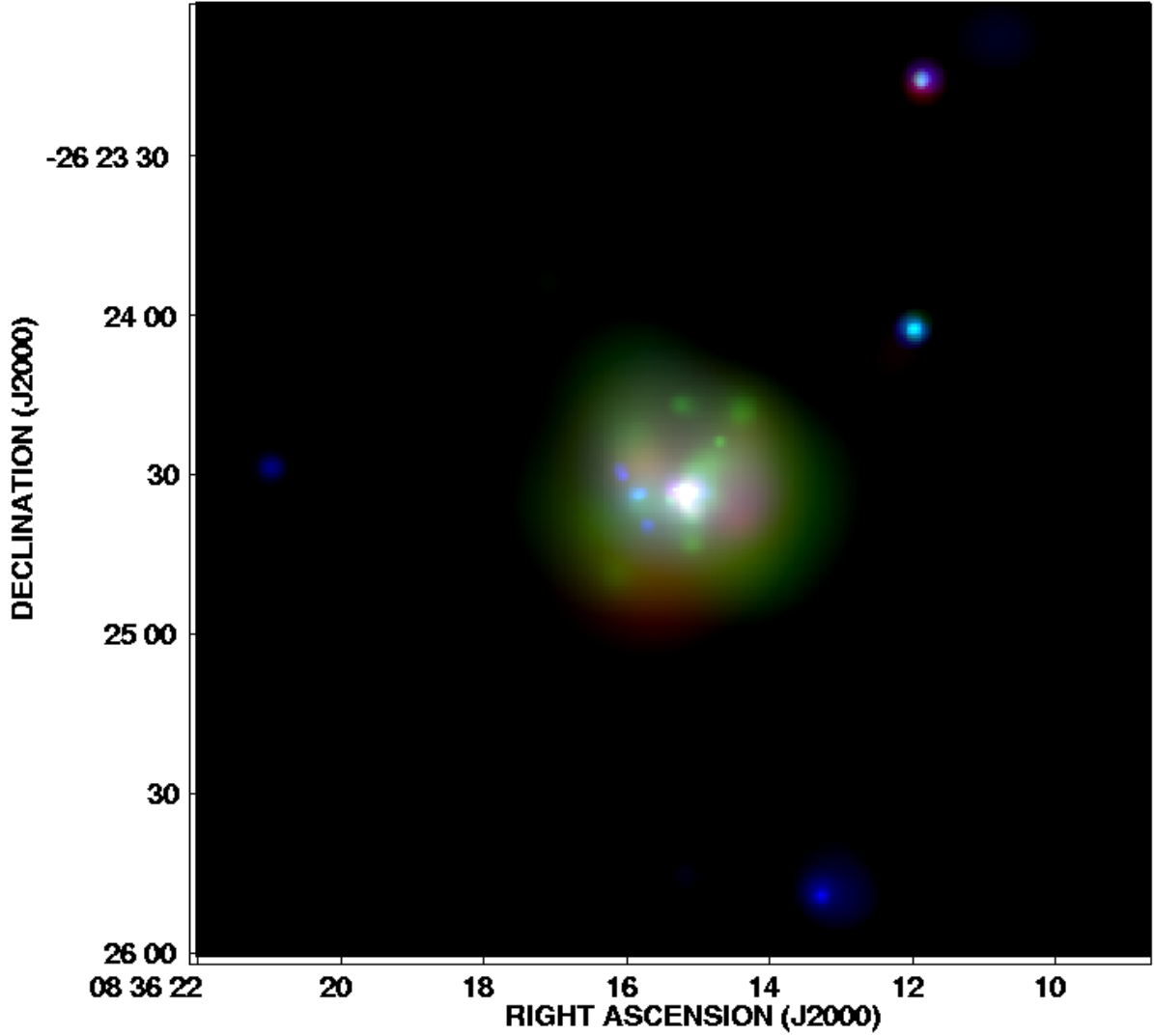}
\caption{Composite 3-color X-ray image of Henize~2-10 with the soft 0.3--0.7 keV 
emission in red,
the medium 0.7--1.1 keV emission in green, and the hard 1.1--6.0 keV emission in blue.   }
\label{3colorX} 
\end{figure}

Figure~\ref{3colorX} show a composite \cxo\ X-ray image of Henize
2-10.  Red represents the soft band 0.3--0.7 keV emission, green
represents the medium band 0.7--1.1 keV, and blue represents the hard
band 1.1--6.0 keV emission.  The two probable X-ray binaries to the
east of the nucleus (sources 2 and 3) appear blue.  The central few
arcseconds of the nucleus have the highest surface brightness and
appear white, denoting a contribution of X-ray emission in all bands.
To the north of the nucleus lies a cone-shaped region with a low X-ray
surface brightness, rimmed on both sides by green emission.  Several
arcseconds to the west of the nucleus lies a red-colored region
indicative of soft diffuse emission with lower foreground extinction.
A similar red-dominated region lies to the
south of the nucleus. 

\begin{figure}
\plotone{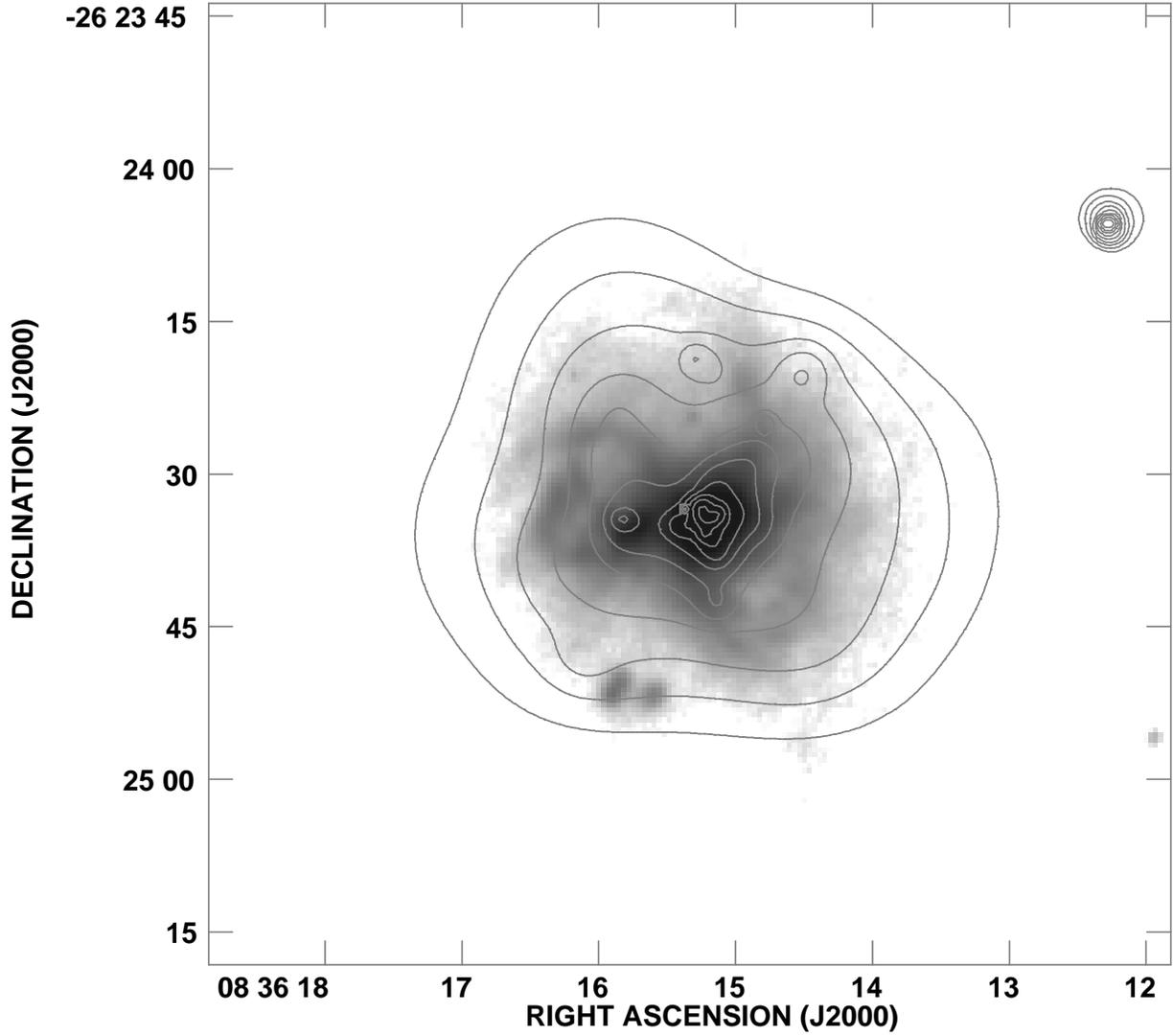}
\caption{$H\alpha$ image of Henize~2-10 (grayscale) with medium band 0.7-1.1 keV
X-rays in contours.  Note the cone-shaped minimum to the north of the
nucleus seen both in X-rays and in the warm ionized gas.  H$\alpha$ shells extend to the
east and southwest.}
\label{Ha-M} 
\end{figure}

Figure~\ref{Ha-M} compares the distribution of the warm ionized medium
(grayscale), as traced by $H\alpha$ emission, overlaid with the medium-band
0.7--1.1 keV X-rays (contours).  Within the inner 10\arcsec, the
X-ray and H$\alpha$ emission is elongated in the NW-SE direction,
similar to the distribution of molecular gas \citep{kobulnicky1995}.
At larger radii, the diffuse X-rays and the H$\alpha$ have similar
morphologies.  Ionized features such as the $H\alpha$ shell to the
east of the nucleus appear as enhancements in the X-ray image as well.
The overall morphology of the galaxy at large radii is nearly
circular, without obvious cones of emission or X-shaped features seen
emanating from the disks of starburst galaxies \citep{strickland2000}.

\begin{figure}
\plotone{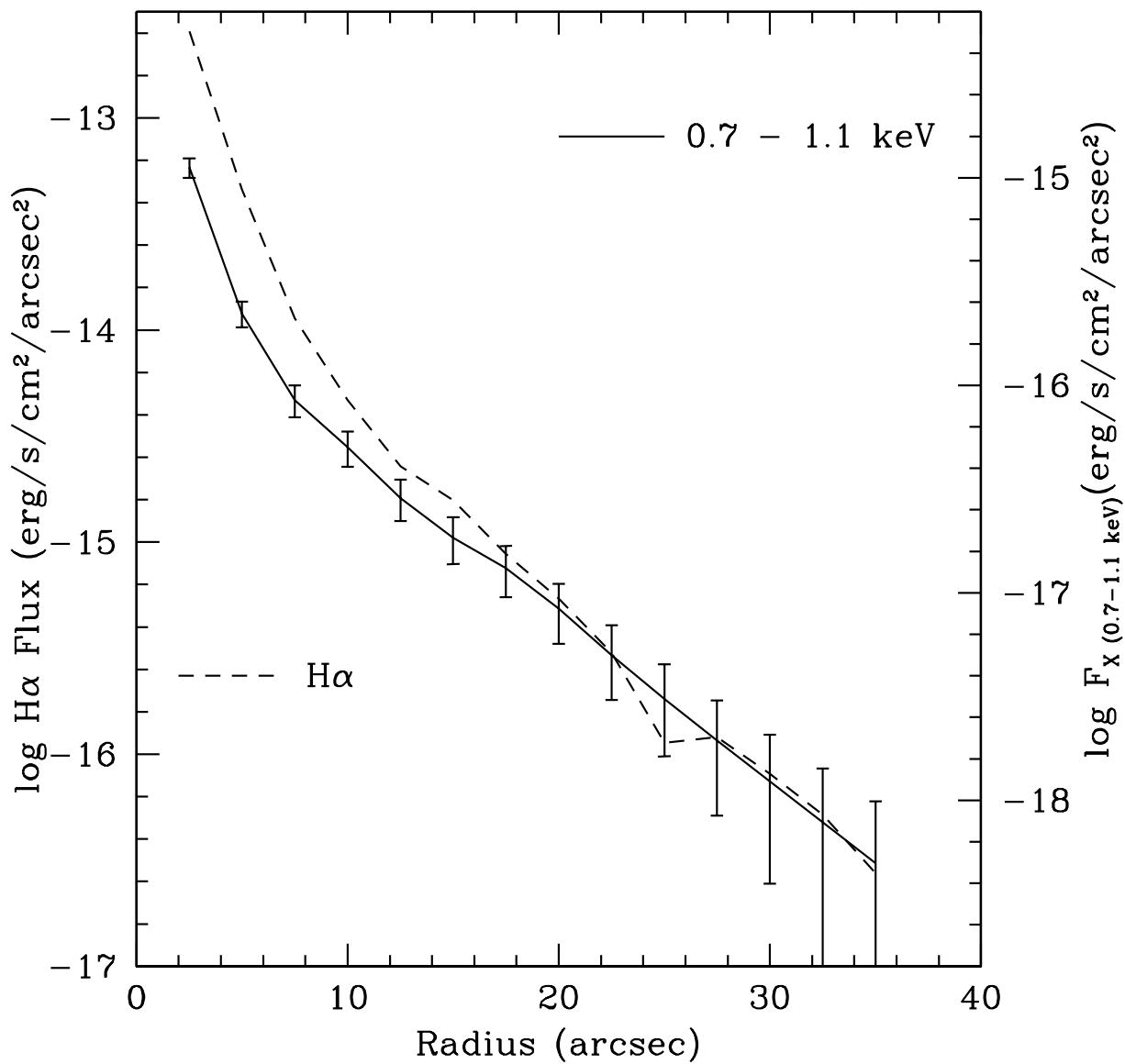}
\caption{Azimuthally averaged radial profile of H$\alpha$ (solid line) and
medium band X-ray (dotted line) surface brightness in Henize~2-10. 
The left abscissa shows the H$\alpha$ scale and the 
right abscissa shows the X-ray scale.  The shape of the radial profiles
in X-rays and H$\alpha$ are nearly identical except within $\sim$10\arcsec\ of 
the nucleus where extinction may be a significant factor. }
\label{radial} 
\end{figure}

The surface brightness profiles of the diffuse X-ray emission and
H$\alpha$ trace each other closely, as found in large samples of
star-forming galaxies \citep{grimes2005}.  Figure~\ref{radial} compare
the azimuthally averaged surface brightness of these two ISM
components as a function of radius using 2.5\arcsec\ wide annuli
centered on the nucleus.  The left abscissa shows the H$\alpha$
surface brightness in units of erg~s$^{-1}$ cm$^{-2}$ arcsec$^{-2}$
and the right abscissa shows the 0.7--1.1 keV X-rays where we have
used the best-fit thermal plasma model (see below) to find that 1
photon s$^{-1}$ corresponds to 3.3$\times10^{-12}$
erg~s$^{-1}$~cm$^{-2}$.  Except for the nuclear region inside
10\arcsec\ where extinction may be a significant factor, these tracers
of the warm and hot phases, respectively, exhibit the same radial
profile.  Inside the 10\arcsec\ nuclear region, the X-ray surface
brightness drops below the H$\alpha$ by a factor of $\sim2.5$ (0.4
dex).  This may be understood as a result of differential extinction
between H$\alpha$, where a column density of 1.5$\times10^{22}$
cm$^{-2}$ corresponds to $\tau~\sim4.6$ ($A_V\sim10$), and the peak of
the thermal X-ray emission at 0.7--1.1 keV where $\tau~\geq5.1$ for
the same foreground column.

\subsubsection{X-rays Relative to the Neutral ISM}

\begin{figure}
\plotone{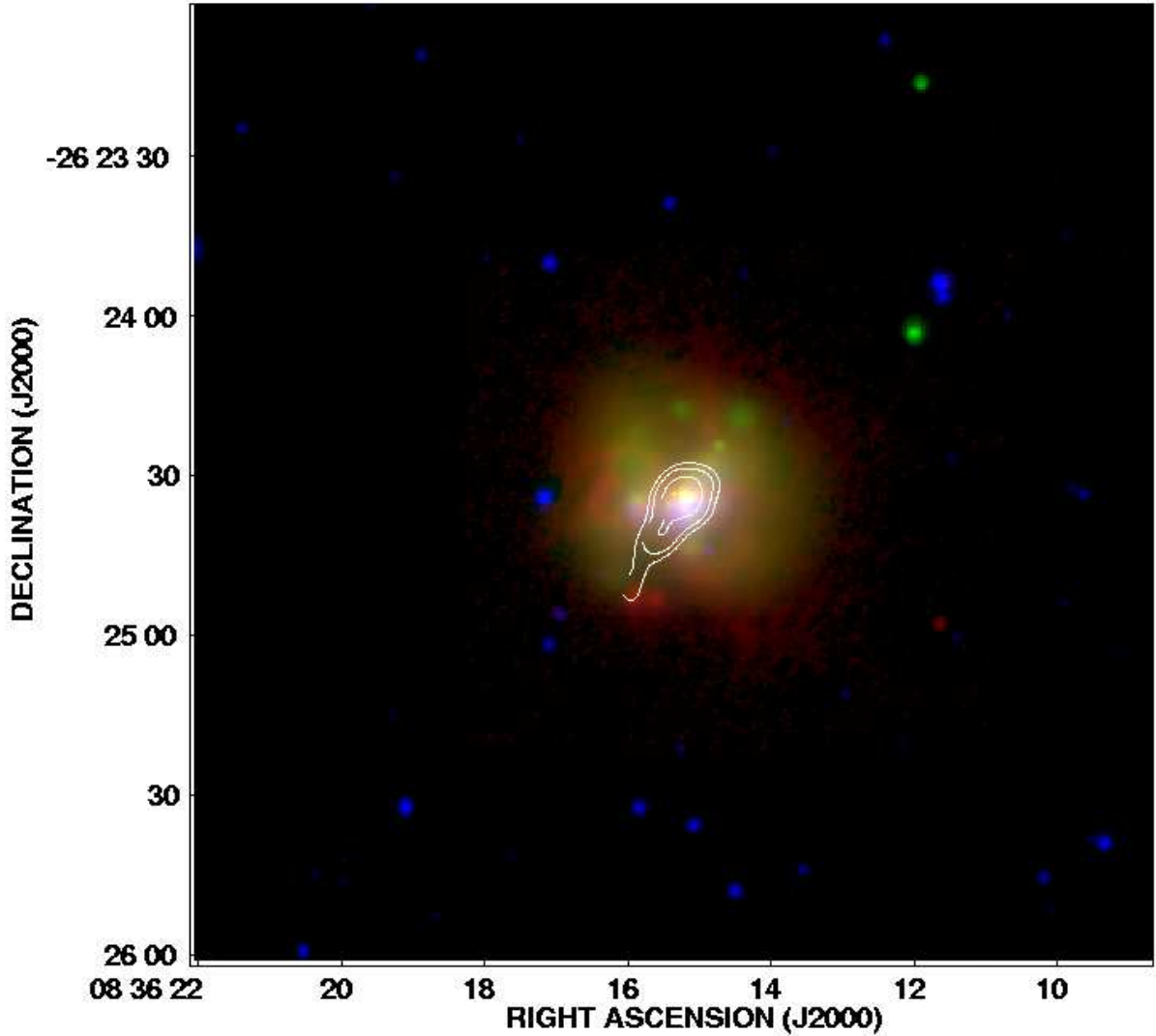}
\caption{Composite three-color image of Henize~2-10 with the \cxo\ 0.7--1.1 keV emission
in green, $H\alpha$ emission in red, and optical 6550 \AA\ continuum image in blue. 
Contours show a putative disk of molecular gas as traced by
$^{12}$CO 1$\rightarrow$0 emission from the Owens Valley Radio Interferometer mapped at a resolution
of 5\arcsec\ \citep{kobulnicky1995}.  $H\alpha$ shells extend to the E--NE and SW, 
perpendicular to the molecular extension.   }
\label{3colorco} 
\end{figure}

The relative distributions of the warm and hot phases are better
pictured in Figure~\ref{3colorco} which shows the $H\alpha$ emission in
red, the 0.7--1.1 keV medium band X-ray emission in green, and the
optical continuum 6550 \AA\ in blue.  Contours show the molecular gas
as traced by $^{12}$CO 1-0 emission observed by the Owens Valley Radio
Interferometer mapped at a resolution of 5\arcsec\ \citep{kobulnicky1995}.  
Although the kinematics of the atomic and molecular gas are
complex, CO velocity-field maps show rotation such that the elongation
in the CO integrated intensity contours probably traces a
molecular ``disk'' within the rotational plane of Henize~2-10.
The $H\alpha$ shells project to the E-NE and SW, presumably 
driven by starburst activity in the nucleus, and indicate a
direction of expansion perpendicular to the molecular disk.

\begin{figure}
\plotone{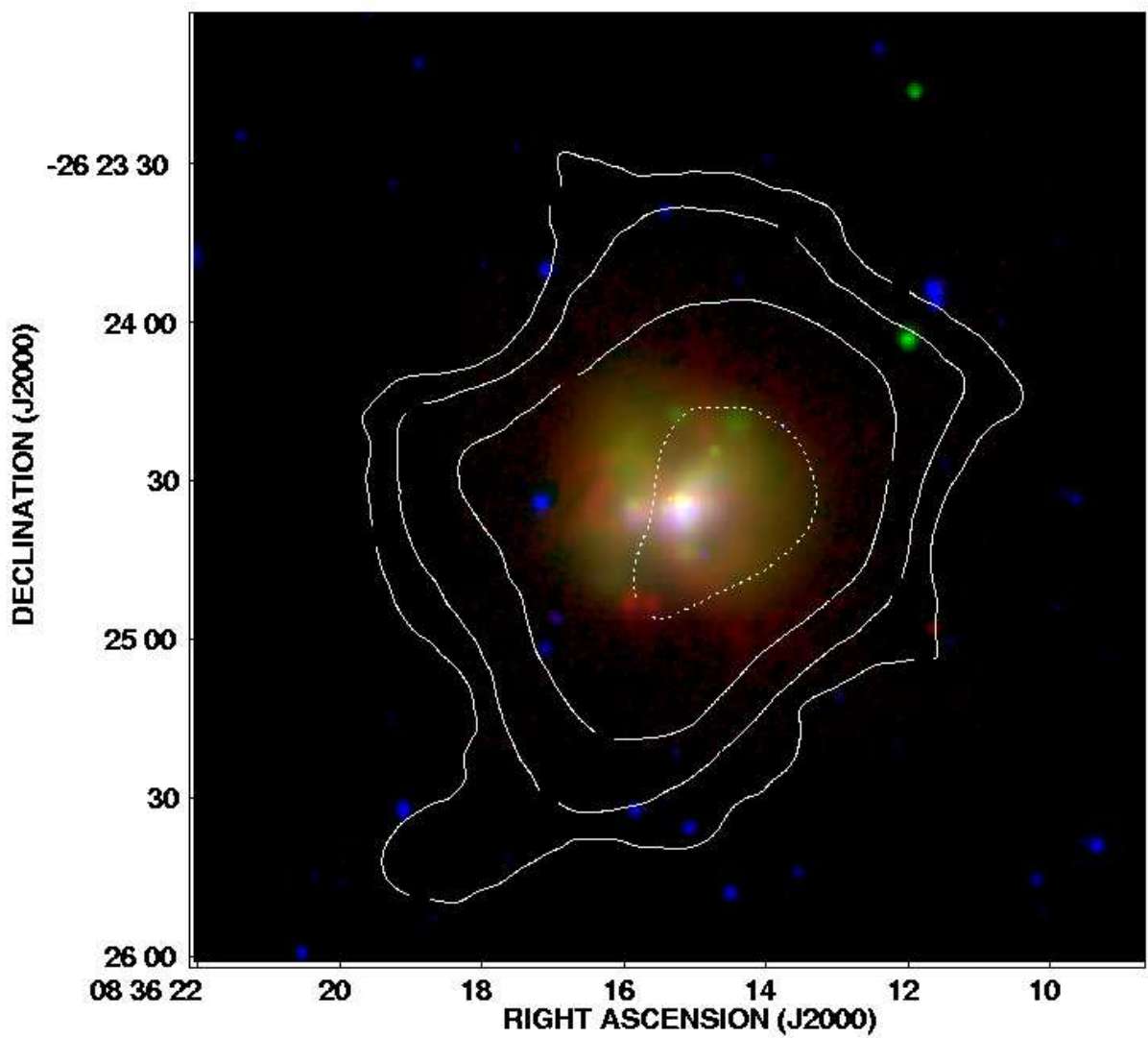}
\caption{Composite three-color image of Henize~2-10 with the \cxo\ 0.7--1.1 keV emission
in green, $H\alpha$ emission in red, and optical 6550 \AA\ continuum image in blue. 
Contours show the neutral atomic gas as traced by
 \ion{H}{1} 21-cm emission from the $VLA$  mapped at a resolution
of 30\arcsec\ \citep{kobulnicky1995}.  Contours correspond to beam-averaged
column densities of $N_{HI}=2\times10^{20},~4\times10^{20},~8\times10^{20},$ and 
$16\times10^{20}$~cm$^{-2}$.  Although the orientation of Henize~2-10 is uncertain,
it appears that the full extent of X-ray and H$\alpha$ emission is
contained within the neutral hydrogen envelope of the galaxy.  }
\label{3colorhi} 
\end{figure}

Figure~\ref{3colorhi} shows the extent of the $H\alpha$ emission
(red), the 0.7--1.1 keV medium band X-ray emission (green), and the
optical continuum 6550 \AA\ (blue) relative to the neutral hydrogen
21-cm emission in contours \citep{kobulnicky1995}.  The contours
correspond to column densities of
N$_{HI}=2\times10^{20},~4\times10^{20},~8\times10^{20},$ and
$16\times10^{20}$~cm$^{-2}$ averaged over a 30\arcsec\ beam.  The
higher-resolution 21-cm and CO maps of \citet{kobulnicky1995} indicate
that Henize~2-10 does contain regions of much larger \ion{H}{1} column
density exceeding $N_{HI}=4\times10^{22}$~cm$^{-2}$ on few arcsecond
scales, but such localized peaks in the gas surface density are not
visible in this low-resolution map.  Although the orientation of
Henize~2-10 is uncertain, Figure~\ref{3colorhi} shows that the full
extent of X-ray and H$\alpha$ emission (at least to the surfce
brightnesses probed in the best existing X-ray and H$\alpha$ observations) is
contained within the $N_{HI}=8\times10^{20}$~cm$^{-2}$ contour.

If the \ion{H}{1} distribution is approximately spherical, then the
starburst-driven bubbles have not penetrated the neutral hydrogen
envelope, at least to the level of 8$\times10^{20}$ cm$^{-2}$.  In
contrast, polar superbubbles would have broken through a thick,
low-inclination \ion{H}{1} disk. The H$\alpha$ emission and neutral
gas absorption kinematics allow that we are seeing an expanding polar
bubble approximately face-on, but do not require
it. \citet{mendez1999} remark that the superbubbles to the east and
north-east are blueshifted, perhaps forming the approaching side of a
polar superbubble. The receding lobe could be largely hidden by
extinction aside from a bit of redshifted H$\alpha$ emission to the
southwest. If the southwest X-ray lobe were viewed through the disk,
the X-ray emission in Figure~7 would appear harder to the southwest
than it does towards the northeast. In fact, however,
Figure~\ref{3colorX} shows that the soft (red) regions are
preferentially in the south and west.  If the lobes are mainly polar,
then the X-ray emission must be coming predominantly from the near
side of the outflow.  Given the evidence for a recent merger in this
galaxy \citep{kobulnicky1995} combined with the ambiguity of the
neutral and ionized gas velocity fields, we conclude that the geometry
of the system defies simple characterization and may be dominated by
transient features generated in the merger and outflow.

\subsubsection{Spectral Fits to the Diffuse ISM}

We extracted a spectrum of the diffuse emission using a annular aperture
centered on the nucleus which exclude the 2\arcsec\ region containing
the nuclear point source. The aperture has inner radius 2\arcsec\ and
outer radius 40\arcsec.  A 30\arcsec\ square region
located 60\arcsec\ north of He~2-10 was used to estimate the
background.  Although spectrum may contain a contribution from faint
unresolved X-ray binaries and the probable X-ray binaries Source 2 and
Source 2, that contribution is small compared to the
diffuse emission.  The diffuse spectrum contains 1040 counts and the
spectrum was binned to produce a minimum of 15 counts per energy bin.

The diffuse emission was modeled with simple X-ray thermal plasma
models in Xspec V11.3.  In all cases we adopted a Galactic foreground
column density of $5\times10^{20}$~cm$^{-2}$ based on the integrated
\ion{H}{1} profile Leiden/Dwingeloo neutral hydrogen survey in this
direction \citep{hartmann1996}.  Table~\ref{spectra.tab} lists the
model description, the best fit parameters for
these models, the reduced $\chi^2$ value as a goodness-of-fit
statistic, and the implied 0.3--6.0 keV unabsorbed luminosity. 

\begin{figure}
\plotfiddle{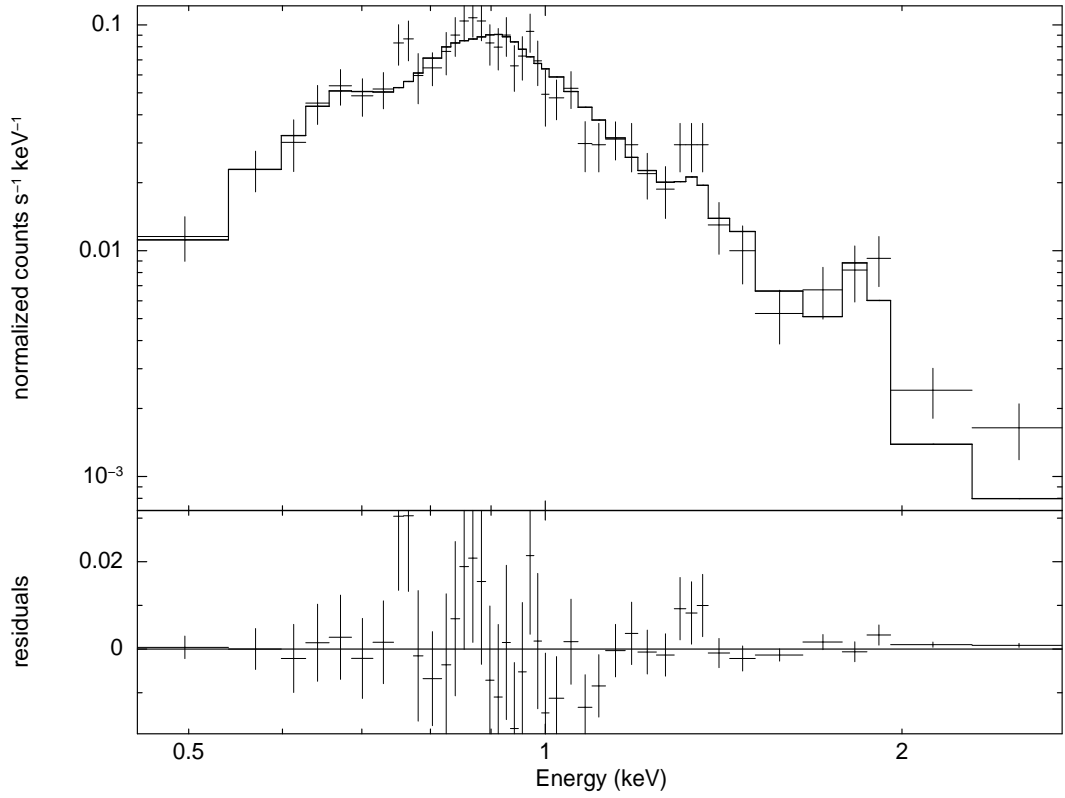}{3.0in}{-90}{300}{400}{0}{0} 
\caption{X-ray spectrum for the diffuse ISM in Henize~2-10 with the best fit 
two-component thermal plasma model (Model 1 of Table~\ref{spectra.tab}) and residuals.     }
\label{spec.model1} 
\end{figure}

Motivated by results in other starburst galaxies where the data
required at least two plasma components with different temperatures
\citep{martin2002, strickland2004, summers2004}, we fit the spectrum
with a two-temperature, solar-abundance model, where each component
has the same absorbing column density.  X-ray bubbles should contain
plasma over a range of temperatures from very hot shocked $\sim$few keV
gas to cooler $\sim$0.1 keV material at the interface of the X-ray and
neutral medium where instabilities and ablation mix the hot gas with
the ambient ISM \citep{strickland2000}.  This model (Model 1 in
Table~\ref{spectra.tab}) produces a good fit to the data
(red. $\chi^2$=0.97) with kT$_1$=0.68 keV and kT$_1$=0.19 keV and
N$_H$=0.5$\times10^{22}$ cm$^{-2}$.  Figure~\ref{spec.model1} shows
the binned data and models spectrum (upper panel ) along with
residuals from the best-fit model (lower panel).  By contrast, a
single-temperature, single-component, solar-abundance thermal plasma
models (Model 2) produces a poor fit to the data (red. $\chi^2$=2.25)
for kT=0.63 keV and N$_H$=0.23$\times10^{22}$ cm$^{-2}$.  Allowing the
metallicity to vary in the single component model (Model 3) produced a
much better fit (red.  $\chi^2$=1.15) for kT=0.63 keV and
N$_H$=0.23$\times10^{22}$ cm$^{-2}$ and $\alpha$/H=0.14 (i.e.,
Z$\simeq$0.14 Z$_\odot$).  However, we reject such a low metallicity
as unphysical given the near-solar abundance of the warm ionized
medium \citep{kkp99}.  Finally, we note that other starburst galaxies
show evidence for enhanced ratios of $\alpha$ to iron-group elements
\citep{martin2002, grimes2005, ott2005}, so we tried a
single-component thermal plasma model where the abundances of $\alpha$
and Fe-peak elements were allowed to vary. The $\alpha$
group elements (C, O, Ne, Mg, Si, S, Ar, Ca) are tied and allowed to
vary independently from the Fe-group elements (N, Al, Fe, Ni for these
purposes).  This model (Model 4) shows an improved fit
(red. $\chi^2$=0.95) over models 2 and 3, having $\alpha$/H=0.78,
$\alpha/Fe$=2.7, kT=0.64 keV, and N$_H$=0.1$\times10^{22}$ cm$^{-2}$.
These parameters are in good agreement with the previous analyses by
\citet{grimes2005, ott2005}.  Both Models 1 and 4 provide good fits to
the data and are consistent with expectations that 1) $\alpha$/Fe
ratios are enhanced in starburst galaxies, and 2) the hot interstellar
medium has a range of temperatures, often parameterized as hot and
cool X-ray components.  Constraining the absolute metallicity of the
hot gas in He~2-10 is not possible with these data because of the
degeneracy between metallicity and the normalization (i.e., $\int n_e
n_H dV /4\pi D^2$) of the X-ray component.
  
In an attempt to identify spatial variations in the X-ray spectral
properties, we extracted spectra using four fan-shaped apertures
centered just outside the nucleus and extending 40\arcsec\ to the
north, south, east, and west, similar to the apertures of
\citet{ott2005}.  All of these apertures exhibited similar spectral
characteristics, with the best fit models showing the same
anti-correlation between best fitting plasma temperature and absorbing
column density. In general, we found that the highest absorbing column
densities are located in the east quadrant, but the limited number of
photons in each spectrum precludes statistically significant
conclusions regarding spatial variations in the diffuse X-ray medium.
On the basis of the color differences in Figure~\ref{3colorX}, we
extracted a spectrum of predominantly red (soft X-ray) regions (small
fan-shaped regions N and S of the nucleus) and compared the best fit
parameters to those for a spectrum of the predominately green (harder
X-ray) regions.  Once again, the small number of photons, particularly
in the red regions of softest X-ray emission, do not constrain the
physical properties of this gas sufficiently well to distinguish
differences from the predominately green spatial regions.

\section{The Outflow in Henize~2-10 }

Regardless of whether an outflow occurring in Henize~2-10 escapes from
the dark matter halo, it is instructive to consider whether the
supernovae in the current burst can drive it.  The $H\alpha$
luminosity of Henize~2-10, measured by \citet{mendez1999} is
$2.1\times10^{41}$ erg~s$^{-1}$ (uncorrected for extinction), leading
to an implied current star formation rate of at least 1.8 \mo\
yr$^{-1}$ \citep{kennicutt1989}.  For a \citet{salpeter1955} initial
mass function with slope $\alpha=-2.35$ between 0.1 \mo\ and 100 \mo,
the number of potential supernovae generated from an instantaneous
burst producing $2.8\times10^{6}$ \mo\ of stars is about 15000 if all
stars over 8 \mo\ end as supernovae.  This is consistent with the 3750
SNRs already present, estimated independently from the nonthermal
radio continuum by \citet{mendez1999} or the 5000 SNRs estimated by
\citet{allen1976}.  The implied current SN rate is one every 250
yrs. The supernova rate is equivalent to a mechanical energy injection
rate varying from few$\times$10$^{39}$ to 10$^{41}$ erg~s$^{-1}$ over
the first 10$^7$ yr, using the online instantaneous burst models of
\citet{leitherer1999}.  The total mechanical energy produced over this
period would be 10$^{55}$ erg.  For continuous star formation models
at a rate of 2 M$_\odot$~yr$^{-1}$, the steady state mechanical energy
injection rate is $\sim$3$\times$10$^{41}$ erg~s$^{-1}$ or about
5$\times10^{55}$~erg over the last 10 Myr.  These should be regarded
as a lower limits, as the mid-IR measurement of the SF rate is nearly
7 M$_\odot$~yr$^{-1}$.  However, this rate is unsustainable for any
extended duration in such a small galaxy.  At this rate all molecular
gas in the galaxy would be consumed in 40 Myr.

We have estimated the mechanical energy contained in the three
H$\alpha$ shells observed to the NE and SW of the nucleus, adopting
the \citet{mendez1999} (projected) expansion speeds of 60 to $\sim$300
\kms.  These shells have a combined H$\alpha$ luminosity of
$\sim$1$\times10^{40}$ erg~s$^{-1}$ based on the fluxes measured on
the H$\alpha$ images of \citet{mendez1999}, scaled to 9.8 Mpc.  For an
assumed density of $n_e=n_H$=100~cm$^{-3}$ and temperature of
10$^4$~K, the case B H$\alpha$ recombination coefficients and
emissivities power of \citet{hummer} lead to a shell mass of
3.5$\times10^5$~M$_\odot$, given a correction factor of 1.4 for
helium.  This implies a shell kinetic energy of $E_{kin}$ =
1.26$\times10^{52} [v_{60}]^2$ erg, where $v_{60}$ is the expansion
speed in units of 60 \kms.  For a bubble expansion speed of 300 \kms,
this implies a maximum kinetic energy of 3.1$\times10^{53}$ erg. Such
kinetic energies would be easily achieved given the magnitude of the
star formation burst estimated above.  Additional thermal energy,
$E_{th}$, contained in the plasma can be estimated by adopting a
volume for the hot gas (29 kpc$^3$), a mean density (0.01 cm$^{-3}$
derived via the X-ray model normalization), and a mean energy per
particle (0.7 keV).  In this manner we find
E$_{th}$=4$\times10^{54}$~erg, i.e., a large fraction (up to 50\%) of
the total energy estimated to have been generated in the recent burst
of star formation, if the lower limit of 2 M$_\odot$~yr$^{-1}$ is
adopted as an average SF rate.  Adopting the larger IR-based SFR leads
to a thermal energy reservoir that is $\sim$15\% of the total energy
generated by recent star formation.

The poorly resolved rotation curve, ambiguous morphology, and
unconstrained inclination of the \ion{H}{1} in Henize~2-10 preclude a
precise estimate of the escape velocity. Using the 21-cm L-V diagram
from \citet{kobulnicky1995}, we estimate a dynamical mass of
$2.94\times10^9$/sin$^2$i \mo\ within a radius of 2100 pc, or a halo
mass of $11.8 \times 10^9$ M$_\odot$ for intermediate inclination, $i
\approx 30$ degrees.  Following \citet{ferrara2000}, we approximate
the dark matter halo with a modified isothermal sphere. The halo scale
radius is $\sim 4.8$~kpc.  In this model, the escape velocity at
1.2~kpc is $\sim 150$ \kms.  Only the highest-velocity component of
the interstellar absorption trough \citep{schwartz2005, schwartz2006}
reveals low-ionization gas exceeding this escape velocity.  The
location of this material along the line-of-sight is unconstrained,
but even if it resides at 2.1~kpc or more, this component represents a
small fraction of the interstellar gas mass. The $H\alpha$ velocities
\citet{mendez1999} also reveal low-ionization gas at speeds exceeding
the escape velocity, but these high speeds are measured about 0.6~kpc
east of knot A.  While some loss of low-ionization material appears
likely, we find no direct evidence that the bulk of the interstellar
gas will escape from Henize~2-10.  This is consistent with the
conclusions of \citet{strickland2000} who found that starbursts were
efficient at expelling metals but not the cool ISM.  If the hotter
plasma component, 0.6--0.8 keV in Table~\ref{spectra.tab}, represents
the post-shock emission from an even hotter wind, then that wind has a
velocity in the 790 to 900 \kms\ range. Since we detect the shock
emission out to 2.1~kpc, then this wind material appears likely to
escape the dark halo. This finding suggests a high fraction of the
metal yield from the starburst is ejected. That Henize~2-10 harbors a
galactic wind is not surprising considering the very high intensity of
star formation; the areal star formation rate exceeds 2
M$_\odot$~yr$^{-1}$~kpc$^{-2}$, which is well above the threshold
suggested for winds \citep{heckman2002}, and may be as much as a
factor of three higher, allowing for extinction.  

\section{Conclusions}

X-ray imaging of Henize~2-10 has revealed a nuclear X-ray source with
a probable luminosity exceeding $10^{40}$ erg s$^{-1}$, suggesting the
presence of an accreting black hole hole within the central 1\arcsec\
(50 pc) with a mass of $>50$ \mo. This source is coincident with a
compact non-thermal radio source that, based on its non-detection in
milliarcsec VLBA data, may be extended over scales of a few tens of
pc.  The angular size, luminosity, and ratio of radio to X-ray fluxes
are consistent with properties of ULX sources and low-luminosity AGN in other
galaxies.

We identify 33 compact X-ray sources in the {\it Chandra} field.
Seven of these appear to be associated with He2-10 and may be luminous
X-ray binaries (2--3) or peaks in the diffuse emission (4).  Seven
additional sources are likely to be foreground Galactic stars while
eighteen sources are probable background AGN.  These latter objects
may constitute good probes of the ISM in and around He~2-10 if their UV
fluxes are sufficiently large for studies with the {\it Hubble Space
Telescope Cosmic Origins Spectrograph}.

Diffuse X-ray emission from Henize~2-10 accounts for 84\% of the X-ray
photons in the 0.3--6.0 keV bandpass, and 98\% of the photons at
intermediate energies (0.7--1.1 keV).  The unabsorbed 0.3--6.0~keV
X-ray luminosity of the diffuse ISM component is $1.0\times10^{40}$
erg s$^{-1}$ for the adopted distance of 9.8 Mpc.  Although the data
are insufficient to place strong constraints on the chemical
properties, the X-ray spectrum is well fit by either a
solar-composition two-temperature plasma with kT$_1$=0.69 keV and
kT$_2$=0.19 keV or a single temperature plasma with $\alpha$/Fe ratios
enhanced to $\sim$2.7 times the solar value.  Both interpretations
have precedents in studies of other starburst galaxies.

We compare, for the first time, the distributions of the cold
molecular, warm atomic, warm ionized, and hot coronal medium in
He~2-10.  The X-ray and H$\alpha$ emission follow a very similar
radial surface brightness profile out to 30\arcsec\ (1.5 kpc). While
both components extend far above and below the nuclear molecular
``disk'', neither is detected at radii beyond the 8$\times10^{20}$
cm$^{-2}$ \ion{H}{1} contour. Unlike larger starburst spiral galaxies,
no prominent outflow cones or extra-planar bubbles are observed beyond
the neutral gas layer.  To the surface brightness levels currently
probed by X-ray and H$\alpha$ observations, the warm ionized and hot
components lie well within the projected \ion{H}{1} envelope.  The
absence of any systematic compilation of X-ray and \ion{H}{1}
distributions in dwarf galaxies makes it unclear whether He~2-10 is
anomalous in this regard.  Evidence for high-velocity outflows in
low-ionization tracers is tenuous, suggesting that it is unlikely that
large fraction of the neutral gas is presently being removed from the
galaxy.  However, the energy of the hot X-ray medium is
E$_{th}$=4$\times10^{54}$~erg.  This constitutes a significant
fraction of the energy estimated to have been generated in the recent
starburst and a reservoir for the continuing development of a global
wind.  The high areal density of star formation can be reconciled with
the lack of prominent outflow signatures if Henize 2-10 is in the very
early stages of developing a galactic wind.

\acknowledgements

We thank Cesar Esteban for providing the H$\alpha$ image of Henize
2-10. We thank Karl Gebhardt for a timely communication about black
holes in low-mass galaxies.  Insightful remarks from an anonmymous
referee improved this manuscript.  H.~A.~K acknowledges support from NASA
through grant NRA-00-01-LTSA-052.  C.~L.~M.  acknowledges support
from NASA through grant G01-2097X, the David and Lucile Packard Foundation,
and from the National Science Foundation under contract 0808161.

{\it Facilities:} \facility{Chandra ()}, \facility{WIYN ()}, \facility{Keck(HIRES)}


\clearpage

\begin{deluxetable}{rrrrrrrrrrrrrrrr}
\rotate
\tabletypesize{\scriptsize} 
\setlength{\tabcolsep}{0.02in} 
\tablewidth{8.7in}
\tablecaption{Henize~2-10 X-ray Point Sources \label{src.tab}}
\tablehead{
\colhead{ID} & 
\colhead{CXO Designation} & 
\colhead{R.A. (2000)} & 
\colhead{Decl. (2000)} & 
\colhead{$C_{tot}$} & 
\colhead{$C_{S}$} & 
\colhead{$C_{M}$} & 
\colhead{$C_{H}$} & 
\colhead{$Rat._{PSF}$} & 
\colhead{R (mag)} & 
\colhead{$\log(f_X/f_V)$} & 
\colhead{Type}   \\ 
\colhead{(1)} & 
\colhead{(2)} & 
\colhead{(3)} & 
\colhead{(4)} & 
\colhead{(5)} & 
\colhead{(6)} & 
\colhead{(7)} & 
\colhead{(8)} & 
\colhead{(9)} & 
\colhead{(10)} & 
\colhead{(11)} & 
\colhead{(12)}   }
\startdata
 1 & CXOU083615.1-262433 & 8 36 15.12 & -26 24 33.7 & 210.0$\pm$15.7  &  13.1$\pm$ 4.2  &  30.5$\pm$  7.1  & 159.4$\pm$ 13.1  &  4.4  &   $>$19 & \nodata & $>$0.2  & XRB/AGN?\\
 2 & CXOU083616.0-262430 & 8 36 16.02 & -26 24 30.6 &  23.6$\pm$ 5.9  &   3.6$\pm$ 2.3  &   4.4$\pm$  3.1  &  13.7$\pm$  4.3  &  2.6  & $>$21.5 & \nodata & $>$0.3  & XRB \\
 3 & CXOU083615.8-262434 & 8 36 15.82 & -26 24 34.1 &  45.1$\pm$ 7.4  &   1.7$\pm$ 1.7  &   7.6$\pm$  3.5  &  36.0$\pm$  6.3  &  3.0  & $>$21.5 & \nodata & $>$0.5  & XRB \\
 4 & CXOU083616.5-262137 & 8 36 16.51 & -26 25 37.6 &   4.9$\pm$ 2.2  &  -0.0$\pm$ 0.0  &   2.0$\pm$  1.4  &   3.0$\pm$  1.7  &  1.0  & $>$21.5 & \nodata & $>-$0.4 & AGN \\
 5 & CXOU083612.3-262405 & 8 36 12.27 & -26 24 05.5 &  52.8$\pm$ 7.3  &   2.0$\pm$ 1.4  &  16.0$\pm$  4.0  &  34.9$\pm$  5.9  &  1.9  &    19.5 &     0.4 &  $-$0.2 & AGN  \\
 6 & CXOU083612.2-262322 & 8 36 12.17 & -26 23 22.2 &  32.2$\pm$ 5.7  &   3.8$\pm$ 2.0  &   7.0$\pm$  2.6  &  21.8$\pm$  4.7  &  1.8  &    19.6 &     0.4 &  $-$0.4 & AGN  \\
 7 & CXOU083600.3-262138 & 8 36 00.34 & -26 21 37.6 & 446.8$\pm$21.3  & 106.8$\pm$10.4  & 203.5$\pm$ 14.3  & 138.1$\pm$ 11.8  &  1.2  &    10.5 &     0.4 &  $-$2.8 & Star \\
 8 & CXOU083625.0-262302 & 8 36 25.04 & -26 23 02.1 &  23.5$\pm$ 4.9  &   8.8$\pm$ 3.0  &  13.0$\pm$  3.6  &   1.9$\pm$  1.4  &  1.3  &    16.4 &     0.4 &  $-1.7$ & Star \\
 9 & CXOU083616.9-261840 & 8 36 16.87 & -26 18 40.3 &  94.9$\pm$10.0  &  29.7$\pm$ 5.5  &  34.7$\pm$  5.9  &  29.5$\pm$  5.6  &  1.3  &    14.9 &     0.4 &  $-1.7$ & Star \\
10 & CXOU083613.5-262543 & 8 36 13.47 & -26 25 43.8 &   9.4$\pm$ 3.2  &  -0.0$\pm$ 0.0  &  -0.0$\pm$  0.0  &   9.8$\pm$  3.2  &  2.2  & $>$21.5 & \nodata &  $>-$0.2 & AGN \\
11 & CXOU083612.2-262018 & 8 36 12.28 & -26 20 18.0 &  17.7$\pm$ 4.4  &   6.9$\pm$ 2.6  &   7.0$\pm$  2.6  &   4.7$\pm$  2.2  &  0.9  & $>$21.5 & \nodata &  $>$ 0.1 & AGN \\
12 & CXOU083610.1-262355 & 8 36 10.12 & -26 23 55.3 &   6.7$\pm$ 2.6  &   2.0$\pm$ 1.4  &   3.0$\pm$  1.7  &   1.9$\pm$  1.4  &  1.6  & $>$21.5 & \nodata & $>-$0.3 & AGN \\
13 & CXOU083609.6-262630 & 8 36 09.66 & -26 26 30.4 &  11.7$\pm$ 3.5  &   2.0$\pm$ 1.4  &   6.0$\pm$  2.4  &   2.9$\pm$  1.7  &  1.5  & 14.3    &     0.4 &  $-$2.9 & Star \\
14 & CXOU083603.0-262715 & 8 36 03.05 & -26 27 15.6 &  25.3$\pm$ 5.1  &   6.6$\pm$ 2.7  &  13.0$\pm$  3.6  &   5.9$\pm$  2.5  &  1.1  & 16.2    &     0.4 &  $-$1.8 & Star \\
15 & CXOU083630.2-262252 & 8 36 30.29 & -26 22 52.7 &   5.4$\pm$ 2.5  &   0.8$\pm$ 1.0  &   1.9$\pm$  1.4  &   2.8$\pm$  1.7  &  1.1  & 12.9    &     0.4 &  $-$3.8 & Star \\
16 & CXOU083630.0-262533 & 8 36 30.13 & -26 25 33.9 &   3.9$\pm$ 2.0  &  -0.0$\pm$ 0.0  &   1.0$\pm$  1.0  &   2.9$\pm$  1.7  &  0.4  & 18.1    &     0.4 &  $-$1.8 & Star \\
17 & CXOU083625.4-262103 & 8 36 25.57 & -26 21 04.6 &   9.0$\pm$ 3.2  &  -0.3$\pm$ 0.2  &   2.8$\pm$  1.7  &   5.7$\pm$  2.5  &  0.9  & $>$21.5 & \nodata & $>-$0.2 & AGN \\
18 & CXOU083614.3-261833 & 8 36 14.40 & -26 18 33.7 &  36.1$\pm$ 6.7  &   4.8$\pm$ 2.5  &  11.6$\pm$  3.5  &  16.8$\pm$  4.3  &  1.2  & $>$21.5 & \nodata & $>$0.4 & AGN \\
19 & CXOU083613.0-261904 & 8 36 13.11 & -26 19 04.4 &  91.6$\pm$ 9.9  &  19.5$\pm$ 4.5  &  20.3$\pm$  4.6  &  47.3$\pm$  6.9  &  1.2  & $>$21.5 & \nodata & $>$0.8 & AGN \\
20 & CXOU083611.7-262133 & 8 36 11.75 & -26 21 33.2 &   3.7$\pm$ 2.0  &   1.0$\pm$ 1.0  &   1.0$\pm$  1.0  &   1.8$\pm$  1.4  &  0.6  & $>$21.5 & \nodata & $>-$0.5 & AGN \\
21 & CXOU083608.2-262035 & 8 36 08.18 & -26 20 35.1 &   6.7$\pm$ 2.6  &  -0.1$\pm$ 0.1  &   3.9$\pm$  2.0  &   2.9$\pm$  1.7  &  0.4  & $>$21.5 & \nodata & $>-$0.3 & AGN \\
22 & CXOU083604.0-261953 & 8 36 03.91 & -26 19 54.6 &   5.6$\pm$ 2.7  &   1.8$\pm$ 1.4  &   2.7$\pm$  1.7  &   1.3$\pm$  1.4  &  0.5  & $>$21.5 & \nodata & $>-$0.4 & AGN \\
23 & CXOU083559.2-262711 & 8 35 59.18 & -26 27 10.8 &   8.6$\pm$ 3.0  &  -0.1$\pm$ 0.1  &   2.9$\pm$  1.7  &   5.9$\pm$  2.5  &  1.0  & $>$21.5 & \nodata & $>-$0.2 & AGN \\
24 & CXOU083552.2-262154 & 8 35 52.17 & -26 21 54.6 &   8.5$\pm$ 3.2  &   0.8$\pm$ 1.0  &   1.9$\pm$  1.4  &   5.0$\pm$  2.5  &  0.4  & $>$21.5 & \nodata & $>-$0.2 & AGN \\
25 & CXOU083626.6-262157 & 8 36 26.59 & -26 21 56.6 &   8.0$\pm$ 3.0  &  -0.3$\pm$ 0.2  &   0.0$\pm$  0.0  &   6.7$\pm$  2.7  &  1.0  & $>$21.5 & \nodata & $>-$0.2 & AGN \\ 
26 & CXOU083602.8-261912 & 8 36 03.10 & -26 19 13.4 &  19.5$\pm$ 4.9  &   2.7$\pm$ 1.8  &   3.7$\pm$  2.0  &  13.1$\pm$  3.8  &  1.0  & $>$21.5 & \nodata & $>$0.2 & AGN \\ 
27 & CXOU083602.1-261850 & 8 36 02.03 & -26 18 50.4 &  14.4$\pm$ 4.4  &   5.4$\pm$ 2.5  &   2.4$\pm$  1.8  &   5.6$\pm$  2.7  &  0.9  & $>$21.5 & \nodata & $>$0.0 & AGN \\ 
28 & CXOU083559.3-261853 & 8 35 59.29 & -26 18 52.6 &   4.4$\pm$ 2.3  &   2.8$\pm$ 1.7  &  -0.2$\pm$  0.2  &   1.0$\pm$  1.0  &  0.6  & $>$21.5 & \nodata & $>-$0.5 & AGN \\
29 & CXOU083559.1-262656 & 8 35 59.11 & -26 26 56.2 &   4.3$\pm$ 2.2  &   0.8$\pm$ 1.0  &   0.9$\pm$  1.0  &   2.8$\pm$  1.7  &  0.6  & $>$21.5 & \nodata & $>-$0.5 & AGN \\ 
30 & CXOU083615.1-262437 & 8 36 15.12 & -26 24 37.8 &   23.1$\pm$8.5  &   0.5$\pm$3.2   &  17.4$\pm$6.8 &   5.9$\pm$4.5 &  1.74  & $>$19 & \nodata &  \nodata & Diffuse   \\
31 & CXOU083615.2-262434 & 8 36 15.22 & -26 24 34.6 &  64.3$\pm$16.9  &  17.8$\pm$6.1   & 56.8$\pm$11.1 &  0.5$\pm$10.5 &  1.94  & $>$19 & \nodata &  \nodata & Diffuse   \\
32 & CXOU083615.3-262433 & 8 36 15.33 & -26 24 33.2 &  22.1$\pm$11.7  &     4$\pm$4.6   &   9.4$\pm$8.1 &   9.6$\pm$7.5 &  1.94  & $>$19 & \nodata &  \nodata & Diffuse   \\
33 & CXOU083615.4-262435 & 8 36 15.40 & -26 24 35.2 &   12.6$\pm$7.7  &     0.5$\pm$3   &   7.9$\pm$6.1 &   0.5$\pm$4.6 &  1.99  & $>$19 & \nodata &  \nodata & Diffuse   \\
\enddata
\tablerefs{
(1) Reference ID \# for this paper;
(2) \cxo\ X-Ray Observatory identifier; 
(3) J2000 Right Ascension from ACIS total band image; 
(4) J2000 Declination; 
(5) Background-subtracted X-ray counts for each source in the 20 ksec observation 
        detected in the 0.3--6.0 keV band and $1\sigma$ uncertainty from photon statistics. The nominal 
	count rate to energy flux conversion is 1 count~s$^{-1}$ = 7.3$\times10^{-12}$ erg~s$^{-1}$~cm$^{-2}$ for
	a Galactic foreground of 1.0$\times$10$^{20}$~cm$^{-2}$.  
(6) Counts in the 0.3--0.7 keV soft X-ray band.   1 count~s$^{-1}$ = 6.6$\times10^{-12}$ erg~s$^{-1}$~cm$^{-2}$. 
(7) Counts in the 0.7--1.1 keV medium X-ray band.  1 count~s$^{-1}$ = 3.3$\times10^{-12}$ erg~s$^{-1}$~cm$^{-2}$.    
(8) Counts in the 1.1--6 keV hard X-ray band.   1 count~s$^{-1}$ = 8.8$\times10^{-12}$ erg~s$^{-1}$~cm$^{-2}$.  
(9) Ratio of source size to PSF size; 
(10) R-band magnitude of optical counterpart, if any, from either our R-band imaging; the 1$\sigma$ uncertainty is typically 0.4 mag.
(11) Log of the ratio of X-ray to optical flux, $f_X/f_V$ : for this
purpose we use $f_X$(erg~s$^{-1}$~cm$^{-2}$) = 5.1$\times10^{-18}$ $\times$ C$_{tot}$ 
where $C_{tot}$ is the number of detected photons from column 5 and 
$f_V(erg~s^{-1}$~cm$^{-2})=1.7\times10^{-9} \times 2.5^{-R}$
where $R$ is the R magnitude from column 10.
(13) Probable source type based on X-ray, radio, optical fluxes or IDs. 
Objects with $log(f_X/f_V)<-0.5$ we tentatively identify as 
stars; Objects with $log(f_X/f_V)>-0.5$ we tentatively identify as 
probable AGN.  Objects in the disk He~2-10 without optical ID
we identify as probable X-ray binaries.  
}
\end{deluxetable}

\begin{deluxetable}{rrrrrc}
\tabletypesize{\scriptsize} 
\setlength{\tabcolsep}{0.02in} 
\tablewidth{4.2in}
\tablecaption{Henize~2-10 X-ray Fluxes and Fractions \label{phot.tab}}
\tablehead{
\colhead{Band (keV)} & 
\colhead{Gross Cts} & 
\colhead{Flux} & 
\colhead{Point Source Cts} & 
\colhead{Net Diffuse Cts}  &
\colhead{Diffuse Fraction}   \\ 
\colhead{(1)} & 
\colhead{(2)} & 
\colhead{(3)} & 
\colhead{(4)} &
\colhead{(5)}  &
\colhead{(6)}  }
\startdata
0.3--6.0 & 1296$\pm$41 &  1.6$\times10^{-13}$  & 202$\pm$17 & 1094$\pm$44 & 84\%$\pm$ 4\% \\
1.1--6.0 &  464$\pm$26 &  5.8$\times10^{-14}$   & 141$\pm$15 &  323$\pm$30 & 69\%$\pm$ 9\% \\
0.7--1.1 &  657$\pm$27 &  8.4$\times10^{-14}$   &  11$\pm$ 8 &  646$\pm$28 & 98\%$\pm$ 4\% \\
0.3--0.7 &  174$\pm$18 &  2.0$\times10^{-14}$   &   9$\pm$ 5 &  165$\pm$19 & 95\%$\pm$11\% \\
\enddata
\tablerefs{
(1) Bandpass;
(2) Background-subtracted counts within 80\arcsec\ diameter aperture; 
(3) Observed flux corresponding to count rate in column 2 in erg~s$^{-1}$~cm$^{-2}$ for best fitting models discussed in Section~3.; 
(4) Total counts in point sources 1, 6, and 7 from Table~\ref{src.tab}; 
(5) Background-subtracted net counts from  diffuse emission ;
(6) Fraction of flux from diffuse component }
\end{deluxetable}

\begin{deluxetable}{clccccccccc}
\tabletypesize{\scriptsize} 
\setlength{\tabcolsep}{0.02in} 
\tablewidth{6.5in}
\tablecaption{Nuclear Source Spectral Fitting \label{spectra_nuc.tab}}
\tablehead{
\colhead{Name} & 
\colhead{Model} & 
\colhead{$N_{H_1}$} & 
\colhead{$N_{H_2}$} & 
\colhead{$kT$}     & 
\colhead{$\Gamma$}  &    
\colhead{Norm$_1$}    & 
\colhead{Norm$_2$}    & 
\colhead{red. $\chi^2$} &   
\colhead{d.o.f.}  &  
\colhead{$L_X$}  \\  
\colhead{(1)} & 
\colhead{(2)} & 
\colhead{(3)} & 
\colhead{(4)} & 
\colhead{(5)} & 
\colhead{(6)} & 
\colhead{(7)} & 
\colhead{(8)} & 
\colhead{(9)} & 
\colhead{(10)} &
\colhead{(11)}   }
\startdata
A & wabs(mekal+wabs*pow) & 1.2$\pm0.2$ & 9.7$\pm1.2$ & 0.4$^\#$ & 4$^\#$ & 1.1$\pm0.2~ \times10^{-4}$ & 24.3$\pm7.6~\times10^{-4}$  & 0.89 & 9 & $1.8\times10^{41}$ \\
B & wabs(mekal+wabs*pow) & 1.0$\pm0.2$ & 4.5$\pm0.9$ & 0.4$^\#$ & 2$^\#$ & 1.1$\pm0.9~ \times10^{-4}$ &  1.0$\pm0.3~\times10^{-4}$  & 1.10 & 9 & $1.1\times10^{40}$ \\
C & wabs(mekal+wabs*pow) & 0.8$\pm0.2$ & 1.9$\pm0.6$ & 0.4$^\#$ & 1$^\#$ & 0.6$\pm0.6~ \times10^{-4}$ &  0.2$\pm0.1~\times10^{-4}$  & 1.33 & 9 & $5.7\times10^{39}$ \\
\enddata
\tablerefs{
(1) Model designation;
(2) Xspec Model formulae;
(3) Foreground absorbing column density for MEKAL component ($10^{22}$~cm$^{-2}$); 
(4) Foreground absorbing column density for power law  component intrinsic to Henize~2-10 ($10^{22}$~cm$^{-2}$); 
(5) Temperature of the MEKAL component (keV)  ; 
(6) Photon index of the power law component   ; 
(8) Normalization of the MEKAL model in units of $10^{-5} \int  10^{-14} n_en_H dV /4\pi D^2$, where $D$ is the
distance to the source; 
(8) Normalization of the power law model   ; 
(9) Reduced $\chi^2$ of fit ;
(10) Degrees of freedom in fit ;
(11) Unabsorbed 0.3--6.0 keV luminosity in erg s$^{-1}$ at a distance of 12.2 Mpc 
for an observed flux of 1.12$\times10^{-11}$~erg~cm$^{-2}$ s$^{-1}$;
(\#)  Denotes parameters held fixed }
\end{deluxetable}

\begin{deluxetable}{clccccccccccccc}
\rotate
\tabletypesize{\scriptsize} 
\setlength{\tabcolsep}{0.02in} 
\tablewidth{8.6in}
\tablecaption{Diffuse Emission Spectral Fitting \label{spectra.tab}}
\tablehead{
\colhead{ID} & 
\colhead{Aperture} & 
\colhead{Model} & 
\colhead{$N_{H_0}$} & 
\colhead{$N_{H_1}$} & 
\colhead{$kT_1$}     & 
\colhead{$\alpha$/H}  &    
\colhead{$\alpha$/Fe}  &    
\colhead{Norm$_1$}    & 
\colhead{$N_{H_2}$} & 
\colhead{$kT_2$}     & 
\colhead{Norm$_2$}    & 
\colhead{red. $\chi^2$} &   
\colhead{d.o.f.}  &  
\colhead{$L_X$}  \\  
\colhead{(1)} & 
\colhead{(2)} & 
\colhead{(3)} & 
\colhead{(4)} & 
\colhead{(5)} & 
\colhead{(6)} & 
\colhead{(7)} & 
\colhead{(8)} & 
\colhead{(9)} & 
\colhead{(10)} & 
\colhead{(11)} & 
\colhead{(12)} & 
\colhead{(13)} &
\colhead{(14)} &
\colhead{(15)}   }
\startdata
1 & 2--40$\prime\prime$ & wabs(wabs(mekal+mekal)) & 0.05$^\#$ & 0.50$\pm$0.03    & 0.0.69$\pm$0.07 & 1$^\#$          & 1$^\#$        &  15.9$\pm$6    & 0.50$\pm$0.02 & 0.19$\pm$0.02  & 8.0$\pm$13   & 0.97 & 38 & $2.2\times10^{40}$ \\
2 & 2--40$\prime\prime$ & wabs(wabs*mekal)        & 0.05$^\#$ & 0.23$\pm$0.04    & 0.63$\pm$0.04   & 1$^\#$          & 1$^\#$        &  12.7$\pm$0.2  & \nodata & \nodata & \nodata                   & 2.25 & 40 & $3.8\times10^{39}$ \\
3 & 2--40$\prime\prime$ & wabs(wabs*mekal)        & 0.05$^\#$ & 0.18$\pm$0.03    & 0.65$\pm$0.04   & 0.14$\pm$0.03   & 1$^\#$        &  45.4$\pm$0.2  & \nodata & \nodata & \nodata                   & 1.15 & 39 & $3.5\times10^{39}$ \\
4 & 2--40$\prime\prime$ & wabs(wabs*vmekal)       & 0.05$^\#$ & 0.10$\pm$0.03    & 0.64$\pm$0.04   & 0.77$\pm$0.4    & 2.7$\pm$0.5   &  20.4$\pm$0.2  & \nodata & \nodata & \nodata                   & 0.95 & 38 & $3.0\times10^{39}$ \\
\enddata
\tablerefs{
(1) Model ID;
(2) Aperture;
(3) Xspec Model formula;
(4) Galactic foreground absorbing column density ($10^{22}$~cm$^{-2}$); 
(5) Foreground absorbing column density for component intrinsic to Henize~2-10 ($10^{22}$~cm$^{-2}$); 
(6) Temperature of the first MEKAL component (keV)  ; 
(7) $\alpha$-element abundance relative to solar  ; 
(8) $\alpha$/Fe abundance ratio relative to solar  ; 
(9) Normalization of the MEKAL model in units of $10^{-5} \int 10^{-14} n_en_H dV /4\pi D^2$ where $D$ is the distance to the source  ; 
(10) Foreground absorbing column density for second component ($10^{22}$~cm$^{-2}$); 
(11) Temperature of the second MEKAL component (keV)  ; 
(12) Normalization of the second MEKAL model in units of $10^{-5} \int  10^{-14} n_en_H dV /4\pi D^2$  ; 
(13) Reduced $\chi^2$ of fit ;
(14) Degrees of freedom in fit ;
(15) Unabsorbed 0.3--6.0 keV Luminosity in erg s$^{-1}$ for a distance of 12.2 Mpc ;
(\#)  Denotes parameters held fixed }
\end{deluxetable}

\begin{figure}
\plotfiddle{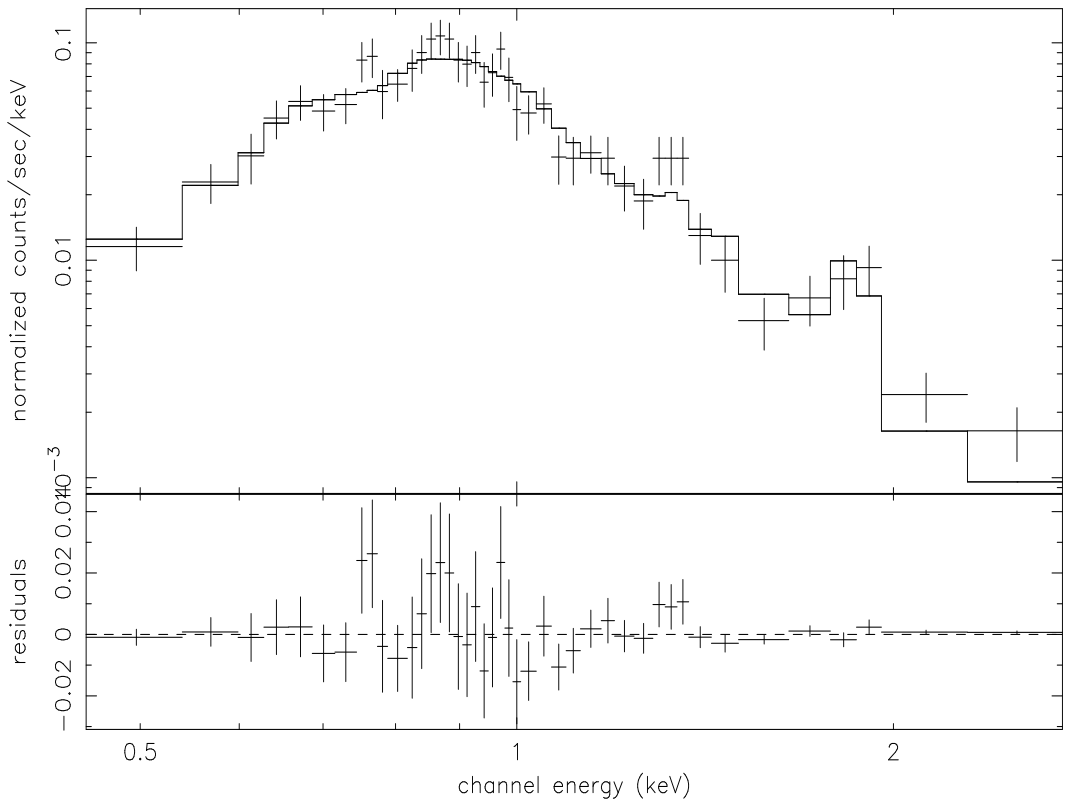}{3.0in}{-90}{300}{400}{0}{0} 
\caption{X-ray spectrum for the diffuse ISM in Henize~2-10 with the best fit 
single-component thermal plasma model with enhanced 
$\alpha$/Fe ratios (Model 4 of Table~\ref{spectra.tab}) and residuals.}
\label{spec.model4} 
\end{figure}

\end{document}